\documentclass[letterpaper,twocolumn,10pt]{article}
\usepackage{usenix2019_v3}

\usepackage{enumitem}
\usepackage{etex}
\usepackage{amsmath,amssymb,amsfonts}
\usepackage{algorithmic}
\usepackage{graphicx}
\expandafter\def\csname ver@subfig.sty\endcsname{}
\usepackage{textcomp}
\usepackage{xcolor}
\usepackage{epsfig,endnotes}
\usepackage{opera}
\usepackage{color}
\usepackage{hyperref}
\usepackage{url}
\usepackage{pgfplots, pgfplotstable}
\usepgfplotslibrary{groupplots}
\usetikzlibrary{patterns}
\usepackage[htt]{hyphenat}
\usepackage{pifont}
\usepackage[linesnumbered,lined,vlined,ruled,commentsnumbered,noend]{algorithm2e}
\SetAlCapNameFnt{\scriptsize}
\SetAlCapFnt{\scriptsize}
\usepackage{tabulary}
\usepackage{booktabs}
\usepackage{bbm}
\usepackage{tikz}
\usetikzlibrary{matrix,positioning}
\usepackage{multirow}
\usepackage{multicol}
\usepackage{placeins}
\usepackage{comment}
\usepackage{mdframed}
\usepackage{soul}
\usepackage{bbding}
\usepackage{colortbl}
\usepackage{pgf-pie}
\usepackage{wasysym}
\usepackage[font={footnotesize}, skip=0pt]{caption}
\def\BibTeX{{\rm B\kern-.05em{\sc i\kern-.025em b}\kern-.08em
   T\kern-.1667em\lower.7ex\hbox{E}\kern-.125emX}}
\usepackage{subfigure}
\usepackage{nopageno}

\newcommand{\cmark}{\ding{51}}%
\newcommand{\xmark}{\ding{55}}%

\usepackage{enumitem}
\setlist{nolistsep,leftmargin=*}

\usepackage[]{authblk}

\renewcommand\Affilfont{\itshape\normalsize}
\makeatletter
\renewcommand\AB@affilsepx{, \protect\Affilfont}
\makeatother

\usepackage{titlesec}
\titlespacing\section{0pt}{10pt plus 2pt minus 2pt}{2pt plus 1pt minus 1pt}
\titlespacing\subsection{0pt}{10pt plus 2pt minus 2pt}{2pt plus 1pt minus 1pt}
\titlespacing\subsubsection{0pt}{10pt plus 2pt minus 2pt}{2pt plus 1pt minus 1pt}

\begin{document}
\newtheorem{mydefinition}{Definition}
\newcommand{\esp}{esp.\@\xspace}
\newcommand{\projectname}{\texttt{CamQuery}\xspace}
\newcommand{\camflow}{\texttt{CamFlow}\xspace}
\newcommand{\system}{\textsc{Sigl}\xspace}
\newcommand{\alacarte}{\emph{\`a la carte}\xspace}
\newcommand{\nec}{\textit{NEC Labs America}\xspace}
\newcommand{\eset}{\texttt{ESET AV Remover}\xspace}
\newcommand{\cL}{{\cal L}}
\renewcommand{\figureautorefname}{Fig.\footnotesize}
\renewcommand{\tableautorefname}{Table}
\renewcommand{\algorithmautorefname}{Alg.}
\renewcommand{\sectionautorefname}{\S}
\renewcommand{\subsectionautorefname}{\S}
\renewcommand{\subsubsectionautorefname}{\S}
\providecommand*{\lstlistingautorefname}{Listing}
\newcommand*\circled[1]{\tikz[baseline=(char.base)]{
            \node[shape=circle,draw,inner sep=0pt] (char) {{\small#1}};}}
            
\date{}

\newmdtheoremenv[
hidealllines=true,
leftline=true,
innertopmargin=0pt,
innerbottommargin=0pt,
linewidth=2pt,
linecolor=gray!40,
innerrightmargin=0pt,
]{definitionii}{Definition}

\newlength{\bibitemsep}\setlength{\bibitemsep}{.2\baselineskip plus .05\baselineskip minus .05\baselineskip}
\newlength{\bibparskip}\setlength{\bibparskip}{0pt}
\let\oldthebibliography\thebibliography
\renewcommand\thebibliography[1]{%
  \oldthebibliography{#1}%
  \setlength{\parskip}{\bibitemsep}%
  \setlength{\itemsep}{\bibparskip}%
}

\title{\system: Securing Software Installations Through Deep Graph Learning\thanks{\system is pronounced as ``seagull''.}}

\author[1]{Xueyuan Han}
\author[2]{Xiao Yu}
\author[3]{Thomas Pasquier}
\author[4]{Ding Li}
\author[2]{Junghwan Rhee}
\author[1]{James Mickens}
\author[5]{Margo Seltzer}
\author[2]{Haifeng Chen}
\affil[1]{Harvard University}
\affil[2]{NEC Laboratories America}
\affil[3]{University of Bristol}
\affil[4]{Peking University}
\affil[5]{University of British Columbia}


\maketitle

\begin{abstract}
Many users implicitly assume that
software can only be exploited
\textit{after} it is installed.
However,
recent supply-chain attacks
demonstrate that application integrity
must be ensured
during installation itself.
We introduce \system,
a new tool for detecting malicious behavior
during software installation.
\system collects traces of system call activity,
building a data provenance graph
that it analyzes using a novel autoencoder architecture with
a graph long short-term memory network (graph LSTM)
for the encoder and a standard multilayer
perceptron for the decoder.
\system flags suspicious installations
as well as the specific installation-time processes
that are likely to be malicious.
Using a test corpus of
625 malicious installers
containing real-world malware,
we demonstrate that \system has
a detection accuracy of 96\%, outperforming
similar systems from industry and academia
by up to 87\% in precision and recall and 45\% in accuracy.
We also demonstrate that \system can pinpoint the
processes most likely to have triggered malicious behavior,
works on different audit platforms and operating systems, and
is robust to training data contamination and adversarial attack.
It can be used with application-specific models, even in the presence
of new software versions, as well as \emph{application-agnostic} meta-models
that encompass a wide range of applications and installers.

\end{abstract}

\section{Introduction}
\label{sec:introduction}
Software installation is risky.
Installer programs
often execute with administrative privileges,
providing installation-time attackers
with powerful capabilities
to immediately corrupt a system
or establish longer-term persistent threats.
Signed installation packages
verify a package's origin,
but not its semantic integrity---installers
can be corrupted before they are signed.
Thus,
as post-installation malware detection
has become more sophisticated,
corruption of digital supply chains
increased by 78\% in the one year from 2018 to 2019~\cite{symantec2019}.
For example,
CCleaner is a popular application
for removing unused files on desktop computers.
In 2017,
attackers breached several workstations
belonging to its developers,
inserting bot software into the official application.
The compromised installer
was downloaded by 2.27 million users, including employees from
major tech companies (\eg Google, and Microsoft) 
before being detected and removed~\cite{swaticcleaner}.

Unfortunately,
there are no strong defenses
against malicious installation. 
Fingerprint-based malware detection
is easy to evade
by tweaking a few bytes of installation data~\cite{kapravelos2013revolver}.
Content-agnostic tools
try to blacklist
the untrusted servers and web pages
that host malicious software~\cite{caballero2011measuring}; 
however,
as the CCleaner attack demonstrates,
corrupted supply chains
provide malicious content
via trusted sources.
More sophisticated detection algorithms
assign dynamic reputation scores to file servers~\cite{rahbarinia2016real,stringhini2017marmite}.
However,
calculating reputation scores is difficult,
requiring labeled malware samples~\cite{stringhini2017marmite}
or a priori knowledge about the characteristics of malicious files~\cite{rahbarinia2016real}.

To improve detection accuracy,
server reputation scoring
can be augmented with client-side anomaly detection.
For example,
data provenance frameworks
observe causal interactions
between kernel-level objects,
such as processes, files, and network sockets~\cite{carata2014primer}.
Malicious installers
will manipulate these objects
in ways that are statistically unlikely
(and thus detectable using statistical analysis).
However, approaches
using data provenance~\cite{han2017frappuccino, manzoor2016fast}
are designed for long timescales and unpredictable exploit timings:
a provenance log spans weeks or months of system activity,
with threats potentially arriving
at any moment during the logging period.
To reduce log sizes,
provenance systems
reduce high-fidelity event logs
to lower-fidelity summarizations,
performing intrusion detection on the summaries.
Unfortunately,
summarizations hurt diagnostic ability;
they omit important contextual information
about, for example, the specific processes that malware launched,
and the specific files that malware accessed.
When they correctly detect an anomaly, reconstructing
the low-level details of how the attack unfolded requires
manual work that is difficult and error-prone,
but critical for understanding which attack vectors need to be patched.

\system reduces the manual effort needed to
(1) detect malicious installations and
(2) identify the malicious processes.
We observe that
once a malicious installation begins,
a machine typically exhibits anomalous behavior (\autoref{sec:threat}).
Thus,
\system can afford to collect
high-fidelity (but short-term) provenance graphs,
discarding old ones
if no malicious installations are detected.
\system
analyzes provenance data
using a novel form of \textit{unsupervised} deep learning,
which means that human analysts
do not have to label training sets with both benign and malicious graphs.
Instead,
given a machine which is known to be malware-free,
\system automatically featurizes provenance graphs
using a novel \emph{component}-based 
embedding technique tailored for system graphs (\autoref{sec:framework:embedding}).
It then applies
long short-term memory networks (LSTMs)~\cite{peng2017cross}
to extract the graph features
corresponding to normal behavior.
These features do not rely on any particular malware; 
therefore, they are general and robust against malicious behavior.
When deployed on in-the-wild machines,
\system uses anomaly scores (\autoref{sec:framework:detection})
to calculate how far a machine deviates from the baseline features
(and thus how likely it is
that a machine is experiencing a malicious installation).

We evaluate \system by collecting
baseline data
from an enterprise database 
storing system events from 141 machines at \nec.
Using malicious installers from the wild
(as well as ones that we created ourselves),
we tested \system's
ability to detect malicious installation activity.
\system achieved precision, recall, accuracy, and F-score values
all greater than 0.94;
in contrast, competing systems that we tested
were unable to achieve better than 0.9 on more than a single metric,
producing substantially worse scores on the remaining metrics
(\autoref{eval:results:comparison}).
We also found that \system's ranking system typically produces a small
set of candidate processes responsible for the attack, including
the one actually responsible
(\autoref{eval:results:investigation}).
To demonstrate the applicability and robustness of our approach,
we further evaluate \system on different platforms (\ie Windows and Linux)
and with various adversarial scenarios (\eg data contamination and evasion).

\noindent In summary, we make the following contributions:
\begin{itemize}[leftmargin=*]
	\setlength\itemsep{0em}
    \item We formalize the problem
          of detecting malicious software installation.
          In particular,
          we introduce a new kind of provenance graph,
          called a \textit{software installation graph},
          that records the short-term (but high-fidelity) provenance information
          needed to capture malicious installation activity.
    \item We are the first to apply deep graph learning
          to the automatic detection of
          anomalies in software installation graphs (SIGs).
          Our approach uses
          a novel autoencoder architecture
          layered atop a long short-term memory network.
    \item We present a novel node featurization model for system-level provenance entities
    	   that is generalizable to applications beyond our current project.
    \item We build and thoroughly evaluate \system,
          an unsupervised detection system,
          that identifies malicious installations.
          \system creates SIGs
          using information provided by lightweight audit frameworks
          such as Windows ETW or Linux Audit.
          Thus,
          \system requires no additional infrastructure
          on end hosts,
          besides a daemon
          that collects audit data
          and sends it to a centralized analysis machine.
          \system outperforms current state-of-the-art malware detectors,
          while also providing the unique ability
          to identify the set of processes
          potentially involved
          in malicious installation activity.
    \item To the best of our knowledge, we are the first to investigate graph-based adversarial
            attacks~\cite{wang2019attacking, zugner2018adversarial} \emph{given realistic and practical systems constraints} faced by the attackers.
\end{itemize}

\section{Background \& Motivation}
\label{sec:motivation}

\begin{figure}[t]
	\centering
	\includegraphics[width=\columnwidth]{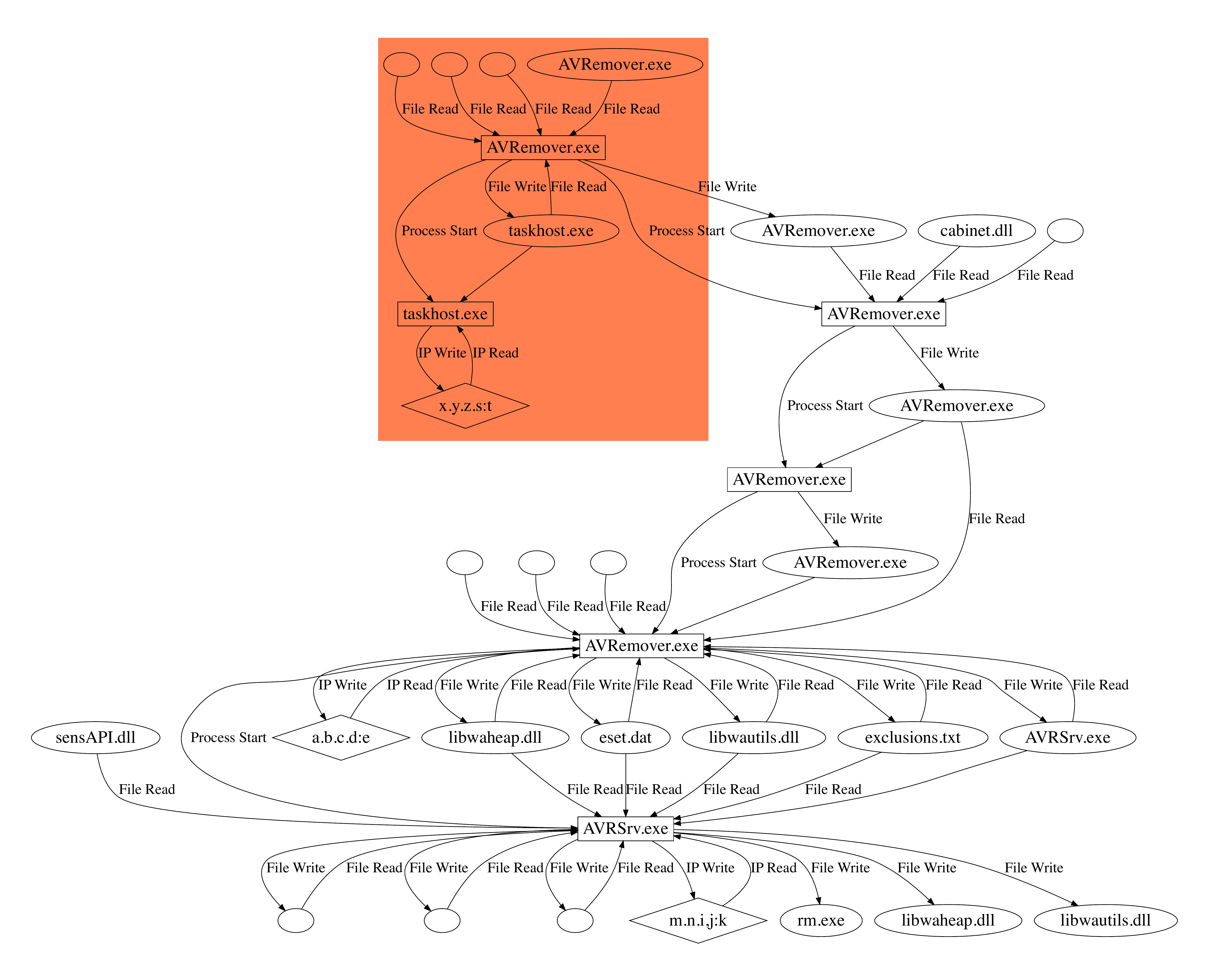}
	\caption{The software installation graph from the attack scenario described in~\autoref{sec:motivation}. The shaded area shows malicious activities not observed in a legitimate installation. We omit some edges, nodes, and node labels for clarity.}
	\label{img:motivation}
	\vspace*{-15pt}
\end{figure}

We simulate 
the following real-world enterprise attack scenario~\cite{scenario}
to illustrate the limitations of existing tools and
motivate {\system}'s design.
Our scenario uses the \texttt{Dharma} ransomware, 
also known as CrySIS, which has become increasingly prevalent in
enterprises~\cite{malwarebytes}.
One important factor that contributes to its popularity 
is its continuous evolution to avoid detection.
We simulate a recent \texttt{Dharma} variant where
the adversary bundles the ransomware tool with a benign anti-virus remover, \eset,
creating a new version of the software package.
The attackers then launch a phishing attack,
impersonating Microsoft, urging enterprise employees to upgrade their anti-virus tool.
When an unsuspecting employee runs the installer,
\texttt{Dharma} runs in the background, encrypting user files,
while the employee interacts with the \eset installer
\footnote{We evaluate \system in this scenario in \autoref{sec:evaluation}.}.
Neither existing malware detection tools nor newer log- or
provenance-based analysis systems are a good match 
for these kinds of attacks because:

\noindgras{Limitations of Malware Detection Tools.}
\label{sec:motivation:limitations}
The \texttt{Dharma} scenario poses several challenges to existing malware detection solutions.
First,
customized variants of \texttt{Dharma} will
effectively evade signature-based malware analysis, including commercial anti-virus detection~\cite{mtrends}.
In fact, many variants of ransomware families, including \texttt{Dharma},
leverage popular installation frameworks (\autoref{sec:eval:data})
to circumvent anti-virus detection without even changing the malware signature~\cite{mcafeeransom}. 
A recent incident demonstrates that,
similar to our motivating scenario,
malware can safely hide in those installation frameworks, 
bypassing all anti-virus products on VirusTotal~\cite{heimdal}.
Second, bundling malicious software with legitimate software
thwarts conventional file reputation analysis~\cite{rahbarinia2016real, stringhini2017marmite}.

Downloader graph analysis~\cite{kwon2015dropper}
or malware distribution infrastructure analysis~\cite{caballero2011measuring}
might have proven effective in this instance if it were possible
to notice the suspicious origin of the bundled installer.
However, if the attackers infiltrated trusted software vendors to distribute
the compromised software package~\cite{webmin}
(\eg the CCleaner incident), 
then, even those approaches would have been rendered ineffective
~\cite{caballero2011measuring}.

In summary, these types of exploits can successfully evade detection from
existing solutions.

\noindgras{Limitations of Log and Provenance Analysis Solutions.}
\label{sec:motivation:solution}
Today's enterprises are rich in commercial threat detection tools and log data;
however, as we show in~\autoref{sec:eval:results},
the log-based commercial TDS~\cite{asi} deployed in our enterprise
produces a large number of false positive alarms,
because it is strict in matching predefined, single-event signatures
(\eg a process should not write to an unknown file).
Newer research prototypes use provenance for intrusion detection~\cite{manzoor2016fast, han2017frappuccino, han2018provenance, pasquier2018ccs},
which provides more contextual analysis, but these systems
value time and space efficiency over fine-grain learning precision.
As such, they tend to 
over-generalize statistical graph features with constrained graph exploration.
For example, \autoref{img:motivation} depicts
the graph structure surrounding the malicious process (\texttt{taskhost.exe}).
Rectangles, ovals, and diamonds represent processes, files, and sockets,
respectively; edges represent relationships between these objects.
The shaded area represents the malicious activity that does not exist in normal \eset installations.
These malicious activities comprise only a small portion of the entire graph,
essentially hiding among the greater number of normal events that take place
during benign installation.
Notice that the graph structure surrounding the malicious process
(\texttt{taskhost.exe})
is similar to that around the benign \texttt{AVRemover.exe}, 
both of which start a new process and communicate with an outside IP address.
Existing IDS cannot distinguish these similar structures, because those systems
use
localized graph analysis (e.g., 1-hop neighborhoods)
that limits their ability to explore more distant relationships
that provide a richer picture of host behavior.
Thus, they produce a large number of false alarms.
Even when the alarms are real, it is difficult to pinpoint the cause of
an alarm, because existing systems 
summarize features, thereby losing details.

These existing systems make rational tradeoffs,
because their goal is whole-system realtime detection over a long time period.
Consequently, they must handle large and fast-growing provenance graphs.
In contrast, \system focuses on the detection of malicious
installation and thus requires a different set of trade-offs.

\noindgras{\system Insight.}
The key insight behind \system is that
\emph{software installation is generally a well-defined, multi-staged process
that can be represented as a bounded, static graph.}
The bounded nature of the graph means that we can analyze the graph in its
entirety rather than having to summarize it.
The multiple stages of installation suggest that we use models that
are inherently temporal.
\system learns both the structure and sequencing of installation
without manual feature engineering.

\section{Problem Formulation and Threat Model}
\label{sec:threat}
We formalize the software installation malware detection problem
as a graph-based outlier detection problem.
Software installation begins when installer execution begins, \eg
the user double clicks on the downloaded package; it terminates when
the installer
process and all its descendants exit.

We characterize the installation behavior of a software package
as a chain of system events leading to its binary files being written to a host system. 
We then define a software installation graph $\mathcal{G}=(V, E)$, an attributed directed acyclic graph (DAG), to represent this event chain.
Nodes $V$ represent system subjects (\ie processes) and objects (\eg files, sockets), 
and edges $E$ record interactions between them.
Given a number of benign installations 
$\mathcal{L} = \{\mathcal{G}^{(s_{1})}, \mathcal{G}^{(s_{2})}, \dots, \mathcal{G}^{(s_{j})}\}$
on endpoint systems $s_{1}, s_{2}, \dots, s_{j}$,
our goal is to learn a model $\mathcal{M}$ of the installation behavior 
that classifies a new installation graph $\mathcal{G}^{(s_{k})}, k \not\in \{1, 2, \dots, j\}$ as benign or malicious.
Given an abnormal $\mathcal{G}$,
we also want to rank process nodes $V_{p} \subset V$ to identify
processes exhibiting the most anomalous behavior.

We assume that the attacker's attempt to infiltrate an enterprise network
through malicious software installation is the initial system breach.
The attacker may distribute malicious installers using phishing emails,
through legitimate software distribution channels (\ie by compromising
the integrity of such channels or acting as a man-in-the-middle),
or by direct access to the network (\ie an insider attack).

\system's threat model assumes the integrity of the underlying OS and audit framework, as is standard for
existing provenance-based systems~\cite{han2017frappuccino, pasquier2018ccs}. 
We further assume the integrity of provenance records,
which can be guaranteed by using existing secure provenance systems~\cite{pasquier2017practical}. 

\section{\system Framework}
\label{sec:framework}
We begin with an overview of \system's architecture and then
present the technical details of each major component.

\subsection{System Overview}
\label{sec:framework:overview}
\begin{figure}[t]
	\centering
	\includegraphics[width=\columnwidth,keepaspectratio,scale=0.8]{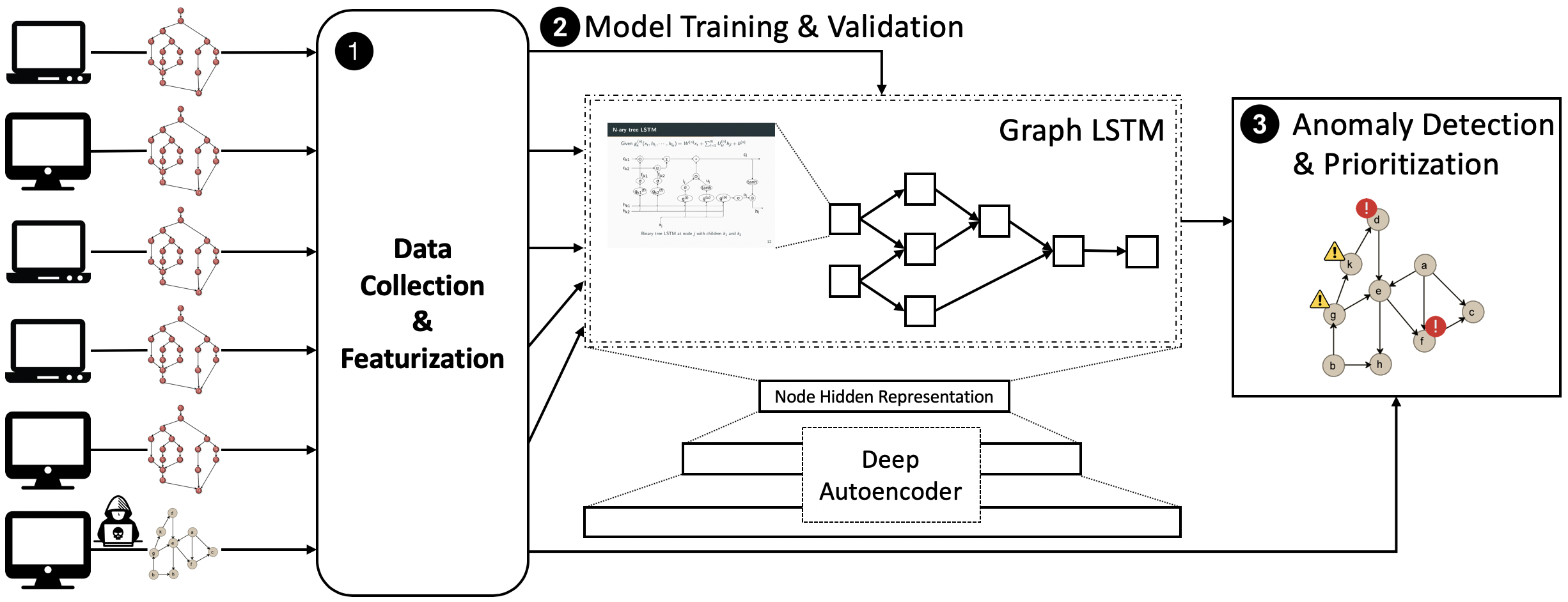}
	\caption{\system collects existing audit data from enterprise workstations and constructs
	software installation graphs to train a deep autoencoder using a graph LSTM as its encoder. 
	The resulting model is used to detect anomalous test graphs and rank nodes within
	the graph based on their anomaly scores.}
	\label{img:overview}
	\vspace*{-5pt}
\end{figure}

\system uses abnormal system behavior to
detect installation of malicious software. 
Its operation consists of three stages:
\circled{1} data collection \& featurization,
\circled{2} model training \& validation, and
\circled{3} anomaly detection \& prioritization.
~\autoref{img:overview} illustrates {\system}'s architecture and workflow.

\noindgras{\circled{1} Data Collection \& Featurization.}
For each software installation considered,
\system gathers audit logs from a collection of machines in the enterprise
and transforms each machine's audit logs into a graphical representation
called a software installation graph (\textit{SIG}, \autoref{sec:framework:graph}).
It then divides the complete set of graphs ($G$) into training
($G_{T}$) and validation ($G_{V}$) sets, with approximately 80\% in the
training set and 20\% in the validation set.
Thus, $\mathcal{G}$ represents a \emph{benign} software installation graph for a particular install.
\system then learns two node embedding models
(\autoref{sec:framework:embedding}) from $G_{T}$.

\noindgras{\circled{2} Model Training \& Validation.}
Given the features learned in \circled{1},
\system trains a deep graph learning model (\autoref{sec:framework:dgl}),
which is a deep autoencoder with a graph
LSTM component as its encoder
and a multilayer perceptron as its decoder.
The autoencoder learns to reconstruct normal process nodes in $\mathcal{G} \in G_{T}$ from
their latent representations encoded by the graph LSTM,
minimizing reconstruction errors.
\system then uses the validation data $G_{V}$ to verify the performance of the learned model
and, using the reconstruction errors, determine the threshold for anomaly
detection.

\noindgras{\circled{3} Anomaly Detection \& Prioritization.}
Given a trained model and threshold (\autoref{sec:framework:detection}),
\system takes audit logs from a new software installation,
generates its corresponding SIG, embeds its nodes using the trained node embedding models,
and uses the autoencoder model to reconstruct all process nodes.
The resulting reconstruction losses are the anomaly scores for each
node.
If the overall anomaly score exceeds the threshold, 
\system classifies the installation as abnormal and
reports a list, sorted by anomaly score, of the most suspicious processes.
System administrators can analyze process behavior through the SIG,
prioritizing the ones with the highest anomaly scores.

\subsection{Software Installation Graphs}
\label{sec:framework:graph}
\begin{table}
\scriptsize
\resizebox{\columnwidth}{!}{%
\begin{tabular}{l l l}
Subject & Object & Event Relationship \\
\hline
\multirow{3}{4em}{\texttt{process}} & \texttt{process} & \texttt{start}; \texttt{end} \\
& \texttt{file} & \texttt{rename}; \texttt{read}; \texttt{write}; \texttt{execute}; \texttt{delete} \\
& \texttt{socket} & \texttt{send}; \texttt{receive}\\
\hline\hline
\end{tabular}}
\caption{System entities and dependency relationships.}
\label{table:graph}
\vspace*{-20pt}
\end{table}

Similar to prior systems~\cite{gehani2012spade,hassannodoze}, \system builds SIGs using common logging frameworks (\eg Windows ETW and Linux Audit) based on standard provenance models~\cite{w3c}.
\system transforms each audit log event into an edge, whose source
represents the subject of the event (\ie the entity responsible for
creating the log record) and whose destination represents the object being
acted upon (\eg files, socket connections).
The edge itself represents a dependency relationship between these entities.
\autoref{table:graph} shows the dependency relationships that we consider in our work.

\system produces the SIG by
backtracking~\cite{king2003backtracking} from the installed software executable(s), represented
as \texttt{file} node(s).
Given a \texttt{file} node, \system adds all edges having that node as their
destination.
It then recursively repeats this procedure for each newly added node,
backtracking to the download of the installation package.
The resulting graph includes
all \texttt{process}es involved in the installation as well as any
\eg dynamically linked libraries (DLL) that were executed.
We apply an adjustable time bound on how far back we track generic
system services (represented as \texttt{process} nodes) that are
commonly invoked during software installation, thereby
minimizing dependency explosion~\cite{lee2013high}.
If the installation produced more than one installed executable,
we combine the backtraces into a single SIG.
As is done in existing provenance based analysis
work~\cite{pasquier2017practical,pasquier2018ccs,milajerdi2019holmes},
we produce acyclic SIGs
by creating multiple node \emph{version}s as the state of the corresponding
subject/object changes~\cite{muniswamy2006provenance}.

\subsection{Node Embedding for System Entities}
\label{sec:framework:embedding}
Machine learning tasks depend on having a
set of informative, discriminative, and independent features~\cite{grover2016node2vec}.
Node featurization is an important building block
in graph 
learning. 

Popular network representation learning frameworks,
such as node2vec~\cite{grover2016node2vec}, 
DeepWalk~\cite{perozzi2014deepwalk}, 
and metapath2vec~\cite{dong2017metapath2vec},
apply natural language processing (NLP) techniques,
most notably word2vec~\cite{mikolov2013distributed},
to derive latent embeddings that capture contextual information
encoded in the networks.
However, these approaches are not designed 
in the context of representing system entities;
in particular, their node features do not encode relationships
between system entities and their functionality within the system,
which are important for downstream graph learning and anomaly detection. 

A good embedding approach for system-level provenance nodes 
must satisfy two important properties.
First, given a system entity that plays a particular role in a system,
its embedding must be close to that of other entities if and only if
their roles are similar. 
For example, 
both system DLLs \path{c:\windows\system32\ntdll.dll}
and \path{c:\windows\system32\kernel32.dll}
contain kernel functions.
Their embeddings should be close to each other in the embedding space
to facilitate downstream graph learning that captures behavioral similarity of 
processes loading and executing these two DLLs.

Second, the embedding approach must generalize to system entities
\emph{not} in the training dataset.
Such entities are especially common in software installation,
because the installation almost always introduces temporary files
and processes that have semi-random path names.
Mishandling such entities (\eg assigning random embeddings)
would cause downstream graph learning to produce
excessive false positives for lack of meaningful features.

We satisfy both of these properties by
featurizing SIG nodes in an embedding space such that
node embeddings encode semantic meanings of the system entities they represent,
while effectively leveraging the classic word2vec~\cite{mikolov2013distributed} learning model.
To the best of our knowledge,
we are the first to use a neural-network-based approach to meaningfully featurize system-level provenance nodes.


\noindgras{Node Embedding in {\system}.} In NLP, 
word2vec embeds words into a low-dimensional continuous vector space,
where words with similar context map closely together.
Given a sequence of words,
word2vec employs a skip-gram model
whose objective is to maximize the log probability of
predicting the context around a given target word.
A fixed size sliding window on the text sequence determines the context.
Assuming the likelihood of observing each context word is independent given the target word,
word2vec maximizes:
\small
$max\sum^{T}_{t=1}logP(w_{t-C}, ..., w_{t+C} | w_{t}) = max\sum^{T}_{t=1}\sum_{-C\leq c\leq C}logP(w_{t+c} | w_{t})$
\normalsize
. $P(w_{t+c} | w_{t})$ is defined by a softmax function:
$
P(w_{t+c} | w_{t}) = \frac{exp({\bf{w}}_{t+c}\cdot{\bf{w}}_{t})}{\sum^{V}_{i=1}exp({\bf{w}}_{i}\cdot{\bf{w}}_{t})}
$
where $C$ is the window size, 
${\bf{w}}_{t+c}$ and ${\bf{w}}_{t}$ are the embeddings of the context word $w_{t+c}$
and the target word $w_{t}$;
$V$ is the vocabulary size.



We apply word2vec as a basis for our embedding approach
to featurize path names associated with SIG nodes. 
Each node in a SIG, whether file, process, or socket, 
corresponds to a file system path name. 
These path names encode important semantic relationships.
Using the same example from earlier,
\path{c:\windows\system32\ntdll.dll}
and \path{c:\windows\system32\kernel32.dll}
reside in the same directory, because they both contain kernel functions.

To map semantically related nodes close in the embedding space, 
we use a \emph{component}-based node embedding model,
where \system learns the embedding of each \emph{component} of a path
and then follows an additive method~\cite{hu2019few} 
to embed a node as the normalized summation of its path components.
\system performs \emph{directed} random walks of fixed length $l$
to construct the causal context for each node:
Given a source node $c_0$ in the SIG,
\system traverses the graph following the direction of the edges.
If a node has more than one outgoing edge,
\system randomly picks an edge to continue the walk.
Let $c_i$ denote the $i^{th}$ node in the walk.
The causal context $\mathcal{C}$ for $c_0$ is $\{c_i | i = 1, \dots, l\}$,
where $c_i$ is generated by the 
distribution:
\small
$P(c_i = v | c_{i-1} = u) = 
	\begin{cases}
		\frac{1}{N} & \text{if $(u, v) \in E$} \\
		0 & \text{otherwise}
	\end{cases}$
\normalsize
, where $N$ is the number of outgoing edges from $c_{i-1}$.
\system generates multiple causal contexts 
for each node.

Unlike existing embedding frameworks~\cite{grover2016node2vec, perozzi2014deepwalk, dong2017metapath2vec},
our approach does not consider each node label as an atomic individual
whose meaning can be derived only from neighboring nodes through random walks along the network;
instead, each path component essentially becomes part of the context.
If we treat the pathname as a single attribute,
such context information is lost in the resulting embedding.


\noindgras{Embedding Unseen Nodes.}
The approach described so far produces embeddings for only those nodes that
have been observed in the training graphs ($G_{T}$).
As mentioned above, software installation often creates temporary folders with meaningless
base path names, sometimes containing machine-specific variations.
In these cases,
\system uses the \alacarte embedding model~\cite{khodak2018carte}, which
follows the distributional
hypothesis~\cite{harris1954distributional}
to efficiently infer the embeddings for out-of-vocabulary (OOV) words via a
linear transformation of additive context embedding
(\ie the average embeddings of context words).
Given the contexts $C_{w}$ of a word $w$ in a vocabulary and
assuming a fixed context window size $|c|$,
a linear transformation is learned through
\small
${\bf{v}}_{w} \approx {\bf{A}}{\bf{v}}_{w}^{additive} = {\bf{A}}(\frac{1}{|C_{w}|}\sum_{c\in C_{w}}\sum_{w'\in c}{{\bf{v}}_{w'}})$
\normalsize
, where ${\bf{v}}_{w}$ are existing high-quality word embeddings.
After learning the matrix ${\bf{A}}$, 
any OOV word $f$ can be embedded in the same semantic space by
\small
${\bf{v}}_{f} = {\bf{A}}{\bf{v}}_{f}^{additive} = {\bf{A}}(\frac{1}{|C_{f}|}\sum_{c\in C_{f}}\sum_{w\in c}{{\bf{v}}_{w}})$
\normalsize
. \alacarte complements the component-based embedding approach,
because it uses the same context-aware and additive mechanism.
Thus, we produce meaningful embeddings using both random walks and
pathname components.
For example,
given an unseen DLL \path{c:\windows\system32\wow64.dll},
our component-based approach allows \alacarte
to take into consideration its parent directories 
(which are the same as those learned for the
\texttt{ntdll.dll} and \texttt{kernel32.dll} nodes),
in addition to any random walks that pass through the node.

\system trains the \alacarte model using $G_{T}$
and uses the trained model to featurize unseen nodes in the validation graphs $G_{V}$ and during live
deployment.

\subsection{Deep Graph Learning on SIGs}
\label{sec:framework:dgl}
	\system uses an autoencoder 
	to learn a robust representation of the process
	nodes in a SIG for both anomaly detection and prioritization.
	The autoencoder consists of two parts: an encoder, for which we use a
	graph long short-term memory network (graph LSTM), and a decoder, for which we
	use a multilayer perceptron (MLP).

	\noindgras{Graph LSTM.}
	An LSTM~\cite{hochreiter1997long}
	captures long-term dependencies of linear sequences.
	Originally developed for NLP tasks, 
	LSTMs have been successfully adapted to a variety of
	sequence modeling and prediction tasks,
	such as 
	program execution~\cite{zaremba2014learning}
	and attack prediction~\cite{shen2018tiresias}.
	The standard LSTM architecture learns
	sequential information propagation only;
	tree-structured LSTMs~\cite{tai2015improved}
	and the more general graph LSTMs~\cite{peng2017cross}
	are two natural extensions that incorporate richer network topologies.
	Graph LSTMs
	allow for flexible graph structures (\eg DAGs) and consider distinct edge types.
	We refer interested readers to Peng \etal~\cite{peng2017cross} 
	for technical details.


	\noindgras{\system's Autoencoder.}
	Intuitively, \system's autoencoder models process nodes as a function
	of those nodes that came before them (temporally) in the SIG.
	The intuition underlying this encoder-decoder architecture 
	is that anomalous nodes are inherently difficult to be represented accurately
	in the embedding space, so trying to reconstruct them produces much larger 
	\emph{reconstruction losses}. \system uses those losses to distinguish abnormal 
	installations from normal ones (\autoref{sec:framework:detection}).

	Although an alternative solution would be to use 
	a binary classifier to determine if a SIG represents a normal
	installation or not, training such a classifier would require more labeled data (both
	normal and anomalous SIGs) than can easily be collected~\cite{axelsson1999base}.
	A set of SIGs dominated by normal installations produces class imbalance,
	and imbalanced two-class training often results in
	poor model performance~\cite{wang2010new}.
	Additionally,
	as an attacker's modus operandi changes over time,
	keeping the trained classifier up-to-date becomes impractical~\cite{shen2018tiresias}.
	Binary classification also provides no insight on the cause of the attack.
	A system administrator would have to manually compare a problematic
	SIG to one or more known good SIGs to identify potentially malicious processes.

	\system's autoencoder addresses limitations of binary classification through
	unsupervised one-class learning that requires only normal SIGs.
	It jointly trains the graph LSTM, as the encoder, with a MLP as the decoder.
	The encoder learns the hidden representation of each process node
	through the graph LSTM,
	taking into account the node's attributes (\ie feature embedding)
	and the hidden representations of all its source nodes (\ie temporality)
	distinguished by the connection types (\ie heterogeneity).
	The decoder then learns to reconstruct the original node embedding from the hidden representation ($h_{j}$).
	The objective is to minimize the reconstruction loss in the training dataset $G_{T}$,
	which consists of only normal SIGs (\ie unsupervised learning).

\subsection{Anomaly Detection}
\label{sec:framework:detection}
The autoencoder's neural network architecture learns to reconstruct process
nodes.
Nodes that show significant topological difference from those encountered
during training
correspond to unexpected changes in installation behavior,
which signals malware activity and
will lead to large reconstruction errors.
\system is a deviation-based anomaly detection system~\cite{an2015variational},
in that it treats process nodes with high reconstruction loss as anomalies.
By ranking process nodes in a SIG by their reconstruction losses (\ie anomaly scores),
\system helps system administrators prioritize analysis of anomalous nodes
and quickly eliminate false alarms.

\system determines a \emph{normality threshold} from the reconstruction losses
observed during validation.
We typically observe that
a small number of process nodes (\eg those with a large number of descendants)
are inherently much more difficult to reconstruct than the rest of the process nodes in a SIG.
These nodes have orders of magnitude higher reconstruction losses.
If we arrange the losses in descending order,
we observe ``natural breaks'' that partition nodes into ranges.
The losses in the first range, 
\ie the ones with the largest values,
represent the ``limits'' of \system's representational capability,
thus providing us with a reasonable baseline to determine the threshold of normal software installation.

\system uses \emph{Jenks' natural breaks}~\cite{jenks1967data}, a statistical mapping method, 
to systematically discover class intervals of the natural breaks in the data series (\ie reconstruction losses).
Jenks' natural breaks is an iterative optimization method
that minimizes intra-class variance while maximizing inter-class variance
by moving one value from the class with the largest deviations from the mean
to the class with the lowest until the sum of the intra-class deviations reaches its minimum~\cite{jiang2013head}.

\change{
\begin{algorithm}[!ht]
	\algsetup{linenosize=\tiny}
	\scriptsize
	\SetAlgoLined
	\DontPrintSemicolon
	\SetKwInOut{Input}{Input}\SetKwInOut{Output}{Output}\SetKwInOut{Variables}{Variables}
	\SetKwData{S}{s}
	\SetKwFunction{GraphAutoEncoder}{GraphAutoEncoder}\SetKwFunction{JenksMaxZoneAvg}{JenksMaxZoneAvg}\SetKwFunction{append}{append}
	\SetKwFunction{standardDeviation}{standardDeviation}\SetKwFunction{mean}{mean}\SetKwFunction{JenksNaturalBreaks}{JenksNaturalBreaks}
	\SetKwFunction{max}{max}
	\Input{Validation graph set $G_{V}$}
	\Output{Normality threshold $\mathcal{T}$}
	\Variables{$thresholdList$ $\leftarrow$ list of largest average losses from $G_{V}$}
	thresholdList $\leftarrow$ [] \;
	\For{$\mathcal{G} \in G_{V}$}{
		$nodeLosses$ = \GraphAutoEncoder{$\mathcal{G}$} \;
		$largestAverageLoss$ = \JenksMaxZoneAvg{nodeLosses} \;
		$thresholdList$.\append{largestAverageLoss} \;
	}
	$std$ $\leftarrow$ \standardDeviation{$thresholdList$} \;
	$mean$ $\leftarrow$ \mean{$thresholdList$} \;
	$\mathcal{T}$ $\leftarrow$ $mean$ + 3 * $std$ \;
	\Return{$\mathcal{T}$}
	
	\SetKwProg{Fn}{Func}{:}{}
	\Fn{\JenksMaxZoneAvg{nodeLosses}}{
		$zone_{1}$, $zone_{2}$, $\dots$ = \JenksNaturalBreaks{nodeLosses} \;
		\Return{\max{\mean{$zone_{1}$}, \mean{$zone_{2}$}, $\dots$}}
	}
	\caption{Normality Threshold}\label{algo:threshold}
\end{algorithm}
}
\vspace*{-10pt}

Using Jenks' natural breaks, which separates reconstruction losses of a SIG's process nodes into multiple ``zones'',
\system identifies the zone with the largest average loss for each validation graph
and constructs a threshold list that contains those average losses for all the validation graphs.
The normality threshold in our experiments (\autoref{sec:evaluation}) is set to be three standard deviations above the average value of the threshold list.
However, 
system administrators can easily adjust this threshold according to their needs (\eg to optimize towards a low false positive/negative rate).
~\autoref{algo:threshold} shows the pseudocode for setting the threshold.
Given the normality threshold,
\system considers any SIG exceeding this threshold as abnormal
and provides system administrators with a list of its process nodes sorted by their anomaly scores.

\section{Evaluation}
\label{sec:evaluation}
\begin{table}[t]
\scriptsize
\centering
\resizebox{\columnwidth}{!}{%
\begin{tabular}{l l l c c c c}
Software Installer & Version & Installation Framework & \# T & \# V & \# BT & \# M \\
\hline\hline
FireFox {\tiny\Asterisk} & 18.1.0 & Mozilla Installer & 86 & 12 & 24 & 20 \\
\hline
FileZilla {\tiny\Asterisk} & 3.35.1 & Nullsoft Scriptable Install System & 88 & 12 & 24 & 40 \\
\hline
PWSafe & 3.48.0 & Nullsoft Scriptable Install System & 88 & 12 & 24 & 40\\
\hline
MP3Gain & 1.2.5 & Nullsoft Scriptable Install System & 88 & 11 & 23 & 40\\
\hline
ShotCut & 18.12.23 & Nullsoft Scriptable Install System & 85 & 12 & 24 & 40\\
\hline 
TeamViewer {\tiny\Asterisk} & 14.4.2669 & Nullsoft Scriptable Install System & 84 & 12 & 24 & 40\\
\hline
Foobar & 1.4.6 & Nullsoft Scriptable Install System & 85 & 12 & 24 & 40\\
\hline
7Zip & 18.5.0 & SFX & 88 & 12 & 24 & 40\\
\hline
TurboVNC & 2.1.2 & Inno Setup & 88 & 12 & 24 & 40\\
\hline
WinMerge & 2.14.0 & Inno Setup & 85 & 11 & 23 & 40\\
\hline
Launchy & 2.5 & Inno Setup & 151 & 21 & 42 & 40\\
\hline
Skype {\tiny\Asterisk} & 8.50.0 & Inno Setup & 80 & 11 & 22 & 40\\
\hline
WinRAR & 5.71.0 & SFX & 84 & 12 & 24 & 20\\
\hline
DropBox {\tiny\Asterisk} & 79.4.143 & DropBox Installer & 84 & 11 & 23 & 20\\
\hline
Slack {\tiny\Asterisk} & 4.0.1 & NuGet Package & 84 & 12 & 24 & 20\\
\hline
Flash {\tiny\Asterisk} & 32.0.0.223 & Flash Installer & 84 & 12 & 24 & 20\\
\hline
OneDrive {\tiny\Asterisk} & 19.103.527 & SFX & 84 & 12 & 24 & 20\\
\hline
NotePad++ & 7.7.1 & NotePad Installer & 85 & 11 & 23 & 20\\
\hline
ICBC Anti-Phishing & 1.0.8 & ICBC Installer & 85 & 11 & 23 & 20 \\
\hline
ESET AV Remover {\tiny$\bigstar$} & 1.4.1 & ESET Installer & 75 & 10 & 21 & 20\\
\hline\hline
\end{tabular}
}
\caption*{\textbf{T}: Training  \textbf{V}: Validation  \textbf{BT}: Benign Test  \textbf{M}: Malicious Installer}
\caption{Software installers used in the experiments.
	Popular software installations in the enterprise are marked with {\tiny\Asterisk}.
	The software discussed in \autoref{sec:motivation} is marked with {\tiny$\bigstar$}.
	Malicious installers are included only in the test dataset.}
\label{table:eval:datasets}
\vspace*{-5pt}
\end{table}

We present a number of experiments to evaluate \system
as a behavior-based malware detection system for secure software installation 
on enterprise end-point systems and an experimental testbed.
We focus on the following research questions:

\noindgras{Q1.} What is the performance of \system in detecting malicious software installation, and
how does it compare to existing commercial TDS 
and other anomaly-based detection systems that leverage data provenance? (\autoref{sec:eval:results}, \autoref{eval:results:comparison})

\noindgras{Q2.} Can \system effectively guide cyber-analysts to quickly identify abnormal processes and potential malware?  (\autoref{eval:results:investigation})

\noindgras{Q3.} Can \system be realistically used in an enterprise setting? (\autoref{eval:results:use}, \autoref{eval:results:sensitivity}, \autoref{eval:results:contamination}, \autoref{eval:results:meta}, \autoref{eval:results:perf})

\noindgras{Q4.} How robust is \system against adversarial attackers? (\autoref{eval:results:adversarial})

\noindgras{Q5.} Can \system generalize to a large variety of software packages and different platforms? (\autoref{eval:results:linux})

\subsection{Datasets}
\label{sec:eval:data}
We describe our methodology to collect audit data from benign and malware-infected
software installations from all the workstations at \nec using Windows ETW.
We also generated additional datasets on our Linux testbed using Linux Audit.
All experiments related to the testbed are discussed in~\autoref{eval:results:linux},
while other sections focus on real-world Windows logs from the enterprise.

\noindgras{Benign Data.}
\label{eval:data:benign}
We collected benign data from the enterprise event database
where system administrators store and monitor company-wide system activity.
We constructed software installation graphs (\autoref{sec:framework:graph})
for popular software in the enterprise.
Software versions are consistent across different machines.
Administrators carefully monitor installations 
to ensure their authenticity. 
We installed additional legitimate and popular software packages ~\cite{filehippo} to increase the size of our dataset.
We also included benign versions of malicious installers found in the wild (\autoref{table:eval:malinstaller}).
\autoref{table:eval:datasets} shows the complete list of software installers used in our evaluation.

\begin{table}[t]
\scriptsize
\resizebox{\columnwidth}{!}{%
\begin{tabular}{l l l l} 
Installer Name & Malware Signature (MD5) & Malware Type & Malware Family \\ 
\hline\hline
TeamViewer & a2fd7c92f1fb8172095d8864471e622a & Win32/Agent & Trojan \\ 
\hline
TeamViewer & a538439e6406780b30d77219f86eb9fc & Win32/Skeeyah.A!rfn & Trojan \\ 
\hline
ESET AV Remover {\tiny$\bigstar$} & d35fa59ce558fe08955ce0e807ce07d0 & Win32/Wadhrama.A!rsm & Ransomware \\ 
\hline
Flash & ab6cef787f061097cd73925d6663fcd7 & Win32/Banload & TrojanDownloader \\ 
\hline
Flash & 7092d2964964ec02188ecf9f07aefc88 & Win32/Rabased & HackTool \\ 
\hline
Flash & 5a9e6257062d8fd09bc1612cd995b797 & Win32/Offerbox & PUA \\ 
\hline\hline
\end{tabular}
}
\caption{Malicious installers found in the wild. The malware discussed in \autoref{sec:motivation} is marked with {\tiny$\bigstar$}.}
\label{table:eval:malinstaller}
\vspace*{-20pt}
\end{table}

\begin{table}[t]
\scriptsize
\resizebox{\columnwidth}{!}{%
\begin{tabular} {l l l} 
Malware Signature (MD5) & Malware Type & Malware Family \\ 
\hline \hline
03d7a5332fb1be79f189f94747a1720f & Win32/VBInject.AHB!bit & VirTool \\ 
\hline
02c7c46140a30862a7f2f7e91fd976dd & Win32/VBInject.ACM!bit & VirTool \\ 
\hline
1243e2d61686e7685d777fb4032f006a & Win32/CeeInject.ANO!bit & VirTool \\ 
\hline
056a5a6d7e5aa9b6c021595f1d4a5cb0 & Win32/Prepscram & SoftwareBundler \\ 
\hline
0f0b11f5e86117817b3cfa8b48ef2dcd & Win32/Prepscram & SoftwareBundler \\ 
\hline
c649ac255d97bd93eccbbfed3137fbb8 & Win32/Unwaders.C!ml & SoftwareBundler \\ 
\hline
02a06ad99405cb3a5586bd79fbed30f7 & Win32/Fareit.AD!MTB & PasswordStealer \\ 
\hline
1537083e437dde16eadd7abdf33e2751 & Win32/Fareit.AD!MTB & PasswordStealer \\ 
\hline
01abfaac5005f421f38aeb81d109cff1 & Win32/Primarypass.A & PasswordStealer \\ 
\hline
c622e1a51a1621b28e0c77548235957b & Win32/Fareit!rfn & PasswordStealer \\ 
\hline
04e8ce374c5f7f338bd4b0b851d0c056 & Win32/DownloadGuide & PUA \\ 
\hline
c62ced3cb11c6b4c92c7438098a5b315 & Win32/Puwaders.A!ml & PUA \\ 
\hline
73717d5d401a832806f8e07919237702 & Win32/KuaiZip & PUA \\ 
\hline
05339521a09cef5470d2a938186a68e7 & Win32/Adload & TrojanDownloader \\ 
\hline
0e8cce9f5f2ca9c3e33810a2afbbb380 & Win32/Gandcrab.E!MTB & Ransomware \\ 
\hline
0f030516266f9f0d731c2e06704aa5d3 & MSIL/Boilod.C!bit & HackTool \\ 
\hline
0ed7544964d66dc0de3db3e364953346 & Win32/Emotet.A!sms & Trojan \\ 
\hline
c60947549042072745c954f185c5efd5 & Win32/Delpem.A & Trojan \\ 
\hline
02346c8774c1cab9e3ab420a6f5c8424 & Win32/Occamy.C!MTB & Trojan \\ 
\hline
0314a6da893cd0dcb20e3b46ba62d727 & Win32/Occamy.B!bit & Trojan \\ 
\hline \hline
\end{tabular}
}
\caption{Real malware used in the experiments to create malicious installers.}
\label{table:eval:malware}
\vspace*{-20pt}
\end{table}


\noindgras{Malware Data.}
\label{eval:data:malware}
We collected malware data from malicious installers discovered in the wild (\autoref{table:eval:malinstaller}).
We also created more than 600 malicious installers by combining benign software installers in~\autoref{table:eval:datasets}
with real malware from VirusShare. 

\autoref{table:eval:malware} lists the malware samples we used in our evaluation.
We randomly selected malware samples from a wide range of malware families
that exhibit diverse behavior.
For example,
trojan attacks and ransomware typically communicate with a remote server,
while malware of the \emph{PUA} family downloads and installs potentially unwanted applications.

We investigated past real-world security incidents (\eg~\cite{scenario, incident1, incident2})
that involve malicious installers as the entry point to high profile attacks
and observed two general approaches to designing malicious installers:

\noindemph{Bundle malware with legitimate installers.}
The attackers create a ``wrapper installer'' that
simultaneously runs an unmodified benign installer in the foreground 
and malware in the background.
We bundle each legitimate installer with every malware sample in~\autoref{table:eval:malware}
to create malicious installers.

\noindemph{Embed malware in legitimate installers.}
The attackers modify an existing benign installer
and embed malware in it.
The installer executes the malware during installation.
This approach requires us to decompile existing installers
and recompile them with malware.

To construct representative malicious installers,
we select software 
using three popular installation frameworks: 
Nullsoft Scriptable Install System (NSIS), 
Inno Setup, 
and SFX, 
and insert every malware sample in~\autoref{table:eval:malware}.
Those frameworks are popular vehicles to spread malware~\cite{heimdal, mcafeeransom}; 
they are also widely used among popular software installers.
Based on our survey of 1,237 Windows applications hosted on Softpedia, 
over 86\% of the installers use
these three frameworks.

\subsection{Implementation \& Experimental Setup}
\label{eval:results:setup}

We implement \system's data collection and graph generation module in Java 8
so that we can use the existing audit event server deployed in our enterprise, which provides APIs only in Java.
\system's core analytic algorithms, 
including node embedding, modeling, and anomaly detection,
are implemented in Python 3.5 and PyTorch 1.1.0 with the CUDA 9.0 toolkit.
We use the Gensim~\cite{rehurek_lrec} library to generate node embeddings for training graphs
and the Deep Graph Library (DGL)~\cite{dgl} to implement deep graph neural networks on top of PyTorch.

For all experiments,
we partition the benign input data into a training set
(70\%), a validation set (10\%), and a false positive test set (20\%).
\autoref{table:eval:datasets} shows the number of
software installation graphs used for training, validation, and testing.

We parameterize the node context for node embedding with window size 5,
10 random walks, each of length 10, and 128 dimensions.
The same window size is used in \alacarte.
We use the skip-gram training algorithm with negative sampling~\cite{guthrie2006closer}
and run 20 epochs over the corpus.

\system performs unsupervised learning, so we need only benign installers for training.
We train \system's deep graph neural network on a system with 
a NVIDIA GTX 1080 Ti GPU with 12 GiB of memory.
We train the model for 100 epochs
with the training batch size set to 25,
validate model performance after every epoch,
and choose the model that produces the best performance on validation data.

\subsection{\system Experimental Results}
\label{sec:eval:results}
\begin{table}
\scriptsize
\resizebox{\columnwidth}{!}{%
\begin{tabular}{l | c c c c c c}
	Method & Precision & Recall & Accuracy & F-Score & FP Percentage \\
	\hline\hline
	\system & 0.94 & \cellcolor{green!25}0.99 & \cellcolor{green!25}0.96 & \cellcolor{green!25}0.96 & 0.06 \\
	\hline
	Commercial TDS~\cite{asi} & 0.07 & 0.59 & 0.90 & 0.12 & 0.93 \\
	\hline
	StreamSpot~\cite{manzoor2016fast} & \cellcolor{green!25}0.97 & 0.52 & 0.72 & 0.68 & \cellcolor{green!25}0.03 \\
	\hline
	Frappuccino~\cite{han2017frappuccino} & 0.95 & 0.12 & 0.51 & 0.21 & 0.05 \\
	\hline\hline
\end{tabular}
}
\caption{Overall \system experimental results compared to other approaches.}
\label{table:eval:results:overall}
\vspace*{-10pt}
\end{table}

\begin{table}
	\scriptsize
\resizebox{\columnwidth}{!}{%
\begin{tabular}{l | c c c c c }
	Software Installer & Precision & Recall & Accuracy & F-Score\\ 
	\hline\hline
	FireFox & 0.78 & 0.70 & 0.77 & 0.74\\
	\hline
	FileZilla & 0.98 & 1.0 & 0.98 & 0.99\\ 
	\hline
	PWSafe & 0.98 & 1.0 & 0.98 & 0.99 \\ 
	\hline
	MP3Gain & 0.98 & 1.0 & 0.98 & 0.99 \\ 
 	\hline
	ShotCut & 0.98 & 1.0 & 0.98 & 0.99 \\ 
	\hline
	TeamViewer & 0.87 & 1.0 & 0.91 & 0.93 \\ 
	\hline
	Foobar & 1.0 & 1.0 & 1.0 & 1.0 \\ 
	\hline
	7Zip & 0.98 & 1.0 & 0.98 & 0.99 \\ 
	\hline
	TurboVNC & 0.95 & 1.0 & 0.97 & 0.98 \\
	\hline
	WinMerge & 0.98 & 1.0 & 0.98 & 0.99 \\ 
	\hline
	Launchy & 0.8 & 1.0 & 0.88 & 0.89 \\ 
	\hline
	Skype & 1.0 & 1.0 & 1.0 & 1.0 \\ 
	\hline
	WinRAR & 0.95 & 1.0 & 0.98 & 0.98 \\ 
	\hline
	DropBox & 0.91 & 1.0 & 0.95 & 0.95 \\ 
	\hline 
	Slack & 0.91 & 1.0 & 0.95 & 0.95 \\ 
	\hline
	Flash & 1.0 & 1.0 & 1.0 & 1.0 \\ 
	\hline
	OneDrive & 0.74 & 1.0 & 0.84 & 0.85 \\ 
	\hline
	NotePad++ & 1.0 & 1.0 & 1.0 & 1.0 \\ 
	\hline
	ICBC Anti-Phishing & 0.95 & 1.0 & 0.98 & 0.98 \\ 
	\hline
	ESET AV Remover & 0.95 & 1.0 & 0.98 & 0.98 \\
	\hline\hline
\end{tabular}
}
\caption{\system experimental result breakdown for each software installer.}
\label{table:eval:results:breakdown}
\vspace*{-20pt}
\end{table}

We evaluate \system's detection performance on 625 malicious installers across a variety of software packages (\autoref{table:eval:datasets}).
\autoref{table:eval:results:overall} shows that 
\system achieves over 90\% precision, recall, accuracy, and F-score
correctly 
identifying all malicious installers 
in the wild.

\system shares a common characteristic with many anomaly-based detection systems
in that it produces more false positives (FPs) than false negatives (FNs),
as reflected by its higher recall (99\%) than precision (94\%).
However,
precision and recall are well balanced,
meaning that \system does not reduce the number of
FPs by compromising its ability to detect actual malicious installers, as do other anomaly-based detection systems (\autoref{eval:results:comparison}).

\autoref{table:eval:results:breakdown} further details
the experimental results for each installer.
It shows that \system delivers consistent performance over a wide range of software exhibiting vastly different installation behaviors.
We investigate two, \texttt{FireFox} and \texttt{OneDrive},
that have slightly lower precision and recall.
We notice that the installation process of these applications
sometimes includes software updates that are captured in SIGs.
\system has difficulty generalizing both installation and update behavior 
from only a few instances of training graphs,
resulting in lower performance than that of other applications.

\subsection{Comparison Study}
\label{eval:results:comparison}
We compare \system to our in-house commercial TDS~\cite{asi}
and two provenance-based research anomaly detection systems, StreamSpot~\cite{streamspotdata} and Frappuccino~\cite{han2017frappuccino}.
We do not compare \system to other commercial TDS,
because they typically require intelligence service subscriptions and 
customized deployment from external vendors.
Similarly, we exclude comparison to academic systems (such as Mastino~\cite{rahbarinia2016real} and Dropper Effect~\cite{kwon2015dropper}, see \autoref{sec:rw}) that
leverage proprietary information from security vendors that is unavailable to us.
\system enables an enterprise to detect threats using \emph{local, enterprise-wide}
information readily available to system administrators;
additional protection from global 
services (\eg Symantec) is complementary.

We conducted a preliminary experiment to show that our malicious installers
(created using real malware in \autoref{table:eval:malware})
can already significantly reduce the efficacy of commercial anti-virus tools,
even without changing malware signatures.
We upload the original malware samples (\autoref{table:eval:malware}) to VirusTotal, 
which scans the samples and reports the number of anti-virus engines that detect them.
On average, 80.8\% of the engines detect the malware listed in \autoref{table:eval:malware}; the lowest detection rate was 70.0\%.  
Testing on our malicious installers,
VirusTotal reports only 42.4\% on average and the minimum detection rate of
10.8\%.
Therefore,
we do not further compare \system to commercial anti-virus tools,
because their limitations are well documented in the literature~\cite{rahbarinia2016real}. 

We briefly describe each evaluated system and discuss the results
in the remainder of this section. 
\autoref{table:eval:results:overall} summarizes the overall results for all the systems in this study.

\noindgras{Commercial TDS.}
The commercial TDS~\cite{asi} inspects every 
event between a process and a file
and determines its potential to be a threat based on two factors:
A) the familiarity of a file
-- if the TDS has some knowledge of the file in the past (based on the file name in the training data),
then it is less likely to be malicious;
B) the diversity of a process
-- if a process writes to many different files,
then the \texttt{write} event itself is less likely to be malicious,
even if the file is unfamiliar to the TDS.

\noindgras{Frappuccino.}
Frappuccino~\cite{han2017frappuccino} detects program anomalies 
by analyzing whole-system provenance graphs~\cite{pasquier2017practical}.
It explores the graph's local neighborhood structures
using a vertex-centric label propagation algorithm to compare the similarity between two provenance graphs.
Based on the assumption that normal behavior of a program
produces similar provenance graphs when it runs on different host systems,
it clusters normal provenance graphs of many running instances of the program as its model
and detects abnormal program runs when their graphs cannot fit into any existing clusters.
We compare \system to Frappuccino,
because both systems make similar assumptions on
the ability to distinguish abnormality from normalcy using provenance graphs.

\noindgras{StreamSpot.}
StreamSpot~\cite{manzoor2016fast} detects host-system intrusions
based on information flow graphs.
Similar to Frappuccino,
it leverages a clustering-based approach using a similarity function that compares two graphs based on their statistics.
It represents each graph as a vector of local substructure frequencies 
and further approximates the vector using a similarity-preserving hashing scheme.
The hashing scheme reduces the dimensionality of the vector
while preserving discriminatory, principal features that better generalize the learned model.
Since StreamSpot claims to detect \emph{any} anomalies on the host system,
we expect it to identify abnormal installation activity.

\noindgras{Experimental Results.}
\autoref{table:eval:results:overall} shows the overall results for all the baseline systems.
For StreamSpot and Frappuccino,
we use the same experimental setups as described in their respective papers or as implemented in their publicly available code repositories.
We notice that StreamSpot's original implementation analyzes only small local substructures in the graph.
Such a constrained graph exploration tends to make graphs look overly similar to each other,
thus resulting in high FNs and low true positives (TPs).
We reimplement StreamSpot to analyze larger graph neighborhoods. 
We show the reimplementation results (\ie better performance) in~\autoref{table:eval:results:overall}.

We see from~\autoref{table:eval:results:overall} that \system significantly outperforms all baseline systems
in terms of recall, accuracy, and F-score.
It reported only 42 FPs among over 1,000 software installations in three months.
On the contrary,
the commercial TDS produces an overwhelmingly large number of FPs
(9,240 events are considered potential threats during the experiment),
resulting in exceedingly low precision
\footnote{The commercial TDS's performance values are computed on a per-event basis,
rather than a per-graph basis, because it has no notion of causality. To understand an alarm,
however, system administrators typically resort to causal analysis, which requires them to
inspect benign events in addition to the alarm-triggering event.}.
The commercial TDS results are consistent with a recent study 
that shows that many enterprises receive at least 300 alerts per day
with more than 50\% being FPs~\cite{fireeye}.
StreamSpot marginally outperforms \system in precision by only 3\%,
at the expense of a much lower recall (by 47\%).
A low recall is typically a product of low TPs and high FNs.
Both StreamSpot and Frappuccino suffer from low recall
because they have limited graph analytical capability. 
They use a vertex-centric approach to explore local graph neighborhoods,
but such exploration ignores temporal relationships among those substructures
and provides only limited views of graph evolution.
As a result,
they are unable to distinguish malicious installers from benign ones,
producing few FPs (\ie higher precision) but many FNs (\ie lower recall).
Although \system reports slightly more FPs,
we show in~\autoref{eval:results:investigation} that it provides auxiliary information
that allows rapid inspection and dismissal of FPs,
which is absent in both StreamSpot and Frappuccino.
Reducing FPs from the hundreds per day of a typical commercial TDS~\cite{fireeye}
to fewer than one per day is a significant step at mitigating ``alert fatigue''~\cite{hassannodoze}.
Existing techniques, such as whitelisting trusted processes during backtracking,
can further reduce these FPs.
The performance of our StreamSpot reimplementation
demonstrates the importance of incorporating structural information in the analysis.
StreamSpot outperformed Frappuccino,
because Frappuccino is unable to retain just the relevant information;
it overgeneralizes its model with ``noise'' in the dataset.

\system benefits from three important features of graph neural networks.
First,
they effectively filter noise.
\system learns to capture relevant information during training,
a data-oriented approach different from the hashing technique used in StreamSpot.
Second,
they preserve long-term memory. 
\system memorizes the sequential procedure of a software installation
and uses this long-term memory to determine the legitimacy of a process during different stages of the installation.
StreamSpot and Frappuccino consider only ``bag-of-subgraphs'' when analyzing provenance graphs.
Third,
they consider non-linear encoding of graph structures.
Graph structures are contexts that help distinguish normal and abnormal process nodes.
\system learns graph structure via its unique neural network architecture, while
the commercial TDS isolates each event from its broader execution context.

\subsection{Prioritizing Anomalous Processes}
\label{eval:results:investigation}
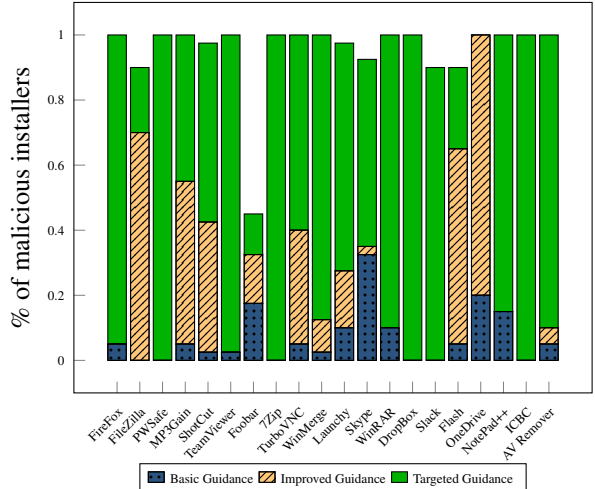
\begin{figure}
\begin{tikzpicture}
	\begin{axis}[
	ybar stacked,
	symbolic x coords={FireFox, FileZilla, PWSafe, MP3Gain, ShotCut, TeamViewer, Foobar, 7Zip, TurboVNC, WinMerge, Launchy, Skype, WinRAR, DropBox, Slack, Flash, OneDrive, NotePad++, ICBC, AV Remover},
	ylabel= \% of malicious installers,
	legend style={at={(0.5,-0.18), font=\tiny},
      	            anchor=north,legend columns=-1},
	x tick label style={rotate=45, anchor=north east},
	xtick=data,
	xtick pos=left,
    	ytick pos=left,
	label style={font=\tiny},
	tick label style={font=\tiny},
	ylabel near ticks,
	xlabel near ticks,
	width=\columnwidth,
	bar width=7,
	height=0.8\columnwidth,
	]
	\addplot[fill={rgb:red,1;green,2;blue,3}, postaction={pattern=dots}] coordinates
		{(FireFox,0.05) (FileZilla,0) (PWSafe,0) (MP3Gain,0.05) (ShotCut,0.025) (TeamViewer,0.025) (Foobar,0.175) (7Zip,0) (TurboVNC, 0.05) (WinMerge, 0.025) (Launchy, 0.1) (Skype, 0.325) (WinRAR, 0.1) (DropBox, 0) (Slack, 0) (Flash, 0.05) (OneDrive, 0.2) (NotePad++, 0.15) (ICBC, 0) (AV Remover, 0.05)};
	\addplot[fill={rgb:orange,1;yellow,2;pink,5}, postaction={pattern=north east lines}] coordinates
		{(FireFox,0) (FileZilla,0.7) (PWSafe,0) (MP3Gain,0.5) (ShotCut,0.4) (TeamViewer,0) (Foobar,0.15) (7Zip,0) (TurboVNC, 0.35) (WinMerge, 0.1) (Launchy, 0.175) (Skype, 0.025) (WinRAR, 0) (DropBox, 0) (Slack, 0) (Flash, 0.6) (OneDrive, 0.8) (NotePad++, 0) (ICBC, 0) (AV Remover, 0.05)};
	\addplot[black, fill=black!30!green] coordinates
		{(FireFox,0.95) (FileZilla,0.2) (PWSafe,1) (MP3Gain,0.45) (ShotCut,0.55) (TeamViewer,0.975) (Foobar,0.125) (7Zip,1) (TurboVNC, 0.6) (WinMerge, 0.875) (Launchy, 0.7) (Skype, 0.575) (WinRAR, 0.9) (DropBox, 1) (Slack, 0.9) (Flash, 0.25) (OneDrive, 0) (NotePad++, 0.85) (ICBC, 1) (AV Remover, 0.9)};
	\legend{Basic Guidance, Improved Guidance, Targeted Guidance}
	\end{axis}
\end{tikzpicture}
\caption{Prioritization of anomalous processes.}
\label{table:eval:results:priority}
\vspace*{-20pt}
\end{figure}

Many existing provenance-based detection systems~\cite{han2017frappuccino, manzoor2016fast, pasquier2018ccs} lack support for postmortem attack investigation,
because their contextual analysis typically requires a holistic understanding of a large provenance (sub)graph.
It is therefore difficult to pinpoint the exact nodes/edges responsible
when a decision is made based on the entire (sub)graph.
Others~\cite{hassannodoze, hossain2017sleuth, milajerdi2019holmes} instead focus on using data provenance to correlate alerts 
from simple edge-based detection systems (\eg commercial TDS) to reduce false alarms and provide attack attribution.
However, they depend on the underlying threat detection system
to reliably report \emph{all} possible threats,
assuming a 100\% detection rate~\cite{hassannodoze}.
\system conducts contextual graph analysis to maintain high detection accuracy.
We show in~\autoref{table:eval:results:priority} that it also assists attack attribution
by accurately identifying anomalous processes within the graph.

We consider three levels of attribution that provide cyber-analysts with increasing degrees of guidance.
We call the malware process (and its associated file) the \emph{target}
and the ranked list generated by \system based on processes' anomaly scores
the \emph{list}.
Note that \system assigns every process and its versions (\autoref{sec:framework:graph}) an anomaly score.
If \system identifies a process among the top 10 in the list that is fewer than 3 hops away from the target (\autoref{table:eval:results:priority}, checks),
we consider \system successfully having provided \emph{basic} guidance.
If the process is ranked among the top 5
and is less than or equal to 3 hops away
(\autoref{table:eval:results:priority}, stripes),
\system has provided \emph{improved} guidance.
Finally, if \system identifies the target among the top 5 in the list
or the target is only 1 hop away from a top-5 process (\autoref{table:eval:results:priority}, solid),
we say that \system offered \emph{targeted} guidance.
These three levels of guidance are based on typical behavior of system administrators, trying to understand the sequence of steps that produced an attack~\cite{king2003backtracking},
and the value (e.g., time savings) that \system brings to the human analysts.

\autoref{table:eval:results:priority} shows that \system is able to 
provide at least basic guidance to identify almost all malicious processes or files for all software installers in the experiment.
In fact,
it provides targeted guidance for at least 10\% of malicious installers in all cases
and more than 50\% of them in the majority (75\%) of the cases.
We investigate two specific examples, \texttt{Foobar} and \texttt{OneDrive},
as they have distinctive results.
\system has difficulty providing effective guidance for about half of the malicious \texttt{Foobar} installers.
We inspected the SIGs of those installers manually
and discovered that \system identifies many versions of a process that originally connects to the malware file as the most anomalous.
It is likely that anomaly scores ``accumulate'' as later versions of the process are being analyzed.
Concrete investigation of how provenance graph versioning affects graph analysis is left for future work.

\system is not able to provide targeted guidance for \texttt{OneDrive},
because \texttt{OneDrive} frequently identifies the update processes in the SIG as among the most anomalous.
As mentioned in~\autoref{sec:eval:results},
a small number of \texttt{OneDrive} training SIGs include both installation and update processes.
\system cannot accurately learn update behavior from only a small number of samples
and therefore incurs high reconstruction losses for those processes.
The same situation is less severe in \texttt{FireFox}, because the update process occurs more frequently in its training data.
However, it does result in lower recall (\autoref{table:eval:results:breakdown}) as the \texttt{FireFox} model attempts to 
generalize both behaviors using a small number of training samples.

Overall,
\system can effectively guide cyber-analysts to quickly identify abnormal processes and potential malware.
Neither StreamSpot nor Frappuccino provides any guidance. 

\subsection{Using \system in an Enterprise}
\label{eval:results:use}
In an enterprise environment,
system administrators configure workstations to include
a standard set of installations. 
When there is a new software release,
the installed software needs to be updated. 
This can lead to a \emph{supply-chain-attack} scenario,
where the attacker exploits a vulnerability in the new release
by compromising the software distribution channel, so 
no legitimate version of the new release is available.
Therefore, we investigate how well \system models generalize across versions,
given that administrators' only defense is the model from the previous
version of the software installation. 


\noindgras{Experimental Setup.}
We installed 
an adjacent version of the software listed in~\autoref{table:eval:datasets}.
In some cases,
our modeled software was already the latest release (at the time of writing);
in those cases, we installed its previous version instead.
To create malicious installers,
we bundle each software installer with a random malware in~\autoref{table:eval:malware}.
\autoref{table:eval:use} lists the versions of the software we use in this experiment.
Note that \texttt{ICBC Anti-Phishing} has only one version.

\newcommand*{\target}{%
   \setbox0=\hbox{\strut}%
   \begin{tikzpicture}
     \filldraw[draw=black,fill=black!30!green] (0,0\ht0) circle[radius=.35em];
   \end{tikzpicture}%
}

\newcommand*{\improved}{%
   \setbox0=\hbox{\strut}%
   \begin{tikzpicture}
     \draw[postaction={pattern=north east lines}, fill={rgb:orange,1;yellow,2;pink,5}] (0,0\ht0) circle[radius=.35em];
   \end{tikzpicture}%
}

\newcommand*{\basic}{%
   \setbox0=\hbox{\strut}%
   \begin{tikzpicture}
     \draw[fill={rgb:red,1;green,2;blue,3}, postaction={pattern=dots}] (0,0\ht0) circle[radius=.35em];
   \end{tikzpicture}%
}

\begin{table}[t]
\scriptsize
\centering
\resizebox{\columnwidth}{!}{%
\begin{tabular}{l l l c c c}
Software Installer & Modeled Version & Test Version & False Alarm & True Alarm & Guidance \\
\hline\hline
FireFox & 18.1.0 & 19.0.1 & \xmark & \cmark & \target \\
\hline
FileZilla & 3.35.1 &  3.34.0 & \xmark & \cmark & \target \\
\hline
PWSafe & 3.48.0 & 3.49.0 & \xmark  & \cmark & \target \\
\hline
MP3Gain & 1.2.5 & 1.2.4 & \xmark  & \cmark & \target \\
\hline
ShotCut & 18.12.23 & 18.12.25 & \xmark  & \cmark & \target \\
\hline 
TeamViewer & 14.4.2669 & 14.5.1691 & \xmark  & \cmark & \target \\
\hline
Foobar & 1.4.6 & 1.5 & \xmark & \cmark & \basic\\
\hline
7Zip & 18.5.0 & 19.0.0 & \xmark & \cmark & \target\\
\hline
TurboVNC & 2.1.2 & 2.2.2 & \xmark & \cmark & \target\\
\hline
WinMerge & 2.14.0 & 2.13.22 & \cmark & \cmark & \target\\
\hline
Launchy & 2.5 & 2.6 & \xmark & \cmark & \target\\
\hline
Skype & 8.50.0 & 8.51.0 & \xmark & \cmark & \basic\\
\hline
WinRAR & 5.71.0 & 5.61.0 & \xmark & \cmark & \improved \\
\hline
DropBox & 79.4.143 & 69.4.102 & \cmark & \cmark & \target\\
\hline
Slack & 4.0.1 & 4.0.2 & \xmark & \cmark & \target \\
\hline
Flash & 32.0.0.223 & 32.0.0.238 & \xmark & \cmark & \improved \\
\hline
OneDrive  & 19.103.527 & 19.086.502 & \cmark & \cmark & \target \\
\hline
NotePad++ & 7.7.1 & 7.7.0 & \xmark & \cmark & \target \\
\hline
ICBC Anti-Phishing & 1.0.8 & N/A & N/A & N/A & N/A \\
\hline
ESET AV Remover & 1.4.1 & 1.3.2 & \xmark & \cmark & \improved\\
\hline\hline
\end{tabular}
}
\caption*{\target: Targeted Guidance  \improved: Improved Guidance  \basic: Basic Guidance}
\caption{Results when testing an adjacent software version on a model.}
\label{table:eval:use}
\vspace*{-20pt}
\end{table}

\noindgras{Experimental Results.}
\autoref{table:eval:use} shows the results for each installer modeled in~\autoref{sec:eval:results}.
We run only one benign and one malicious instance against each model.
If \system considers a benign installer abnormal,
we put a check mark (\cmark) in the \emph{False Alarm} column in~\autoref{table:eval:use};
we check the \emph{True Alarm} column if \system correctly detects a malicious installer.
We see in~\autoref{table:eval:use} that \system continues to maintain high precision and recall
across versions.
Among the 19 benign installers, 
\system correctly classifies 16 of them (84\%) without raising a false positive alarm.
False alerts in our experiments are caused by significant changes in graph structures 
(corresponding to changes in installation behavior) and 
node identities (corresponding to changes in files installed) between two versions.
For example,
\texttt{Dropbox}'s installation behavior changed across the two versions.
We observe that the older version of the \texttt{Dropbox} installer frequently
reads from and executes a temporary file during the installation process.
This behavior creates a large subgraph in the SIG between the file and the process
that is absent in the training dataset.
We quickly identify this difference following the guidance provided by \system.
In~\autoref{sec:discussion},
we further discuss this issue regarding software evolution.
In terms of true alerts,
\system detects all malicious installers with the majority (74\%) having targeted guidance.

\subsection{Sensitivity Analysis}
\label{eval:results:sensitivity}
Anomaly-based detection systems~\cite{chandola2009anomaly} typically
require setting threshold values representing how much of a deviation from
normality constitutes an anomaly.
Thresholds determine the tradeoffs between precision and recall.
Detection systems that are overly sensitive to threshold settings are difficult to use in practice, 
even if there exists an optimal threshold that performs perfect detection.

\system quantifies a normality threshold from the validation dataset based on 
the anomaly scores of individual nodes in the graph (\autoref{sec:framework:detection}).
We demonstrate in~\autoref{fig:eval:sensitivity} that
the anomaly scores of benign and malicious graphs are well-separated with considerable margins
such that \system's detection performance generally does not depend on finding
a precise threshold.

\autoref{fig:eval:sensitivity} shows the average (circled mark), minimum, and maximum (two ends of the error bar)
anomaly scores for benign (blue) and malicious (red) installers for each experiment.  
None of the installs have overlapping benign and malicious ranges,
although the precise break between the ranges is, in fact, installer specific.
However, many of the benign installers have scores orders of magnitude smaller than those of the malicious installers.
For example,
compared to the malicious \texttt{NotePad++} installer with the smallest anomaly score (\autoref{fig:eval:sensitivity}),
even the benign installer with the largest score has a value two orders of magnitude smaller.
Such liberal margins not only make it practical to set anomaly thresholds
but also indicate the likelihood of an installer being benign/malicious.

\pgfplotstableread{
x              y              x-min              x-max
0.004599199	FireFox	0.001641432	0.000811521
4.08E-05	FileZilla	2.54178E-05	0.000145499
1.49E-05	PWSafe	5.99698E-06	3.24764E-05
2.58E-05	MP3Gain	1.33449E-05	1.90639E-05
9.80E-05	ShotCut	7.62345E-05	0.000275961
0.000292584	TeamViewer	0.000175264	0.000188578
0.000320755	Foobar	0.000192146	0.000232903
1.12E-05	7Zip	1.11258E-05	9.23752E-05
8.48E-05	TurboVNC	5.05908E-05	2.22483E-05
0.000209655	WinMerge	3.67902E-05	3.87563E-05
0.000210129	Launchy	0.000157228	0.0001192
0.000488084	Skype	0.00027405	0.000421187
1.93E-05	WinRAR	1.85706E-05	0.000167475
0.000370282	DropBox	0.000117161	0.000273237
0.000408475	Slack	0.000125972	0.000357051
0.000151425	Flash	8.38706E-05	0.000111556
0.000342436	OneDrive	9.30189E-05	0.000188549
1.68E-05	NotePad++	1.20778E-05	0.000106563
4.91E-06	ICBC	2.67007E-06	6.67471E-06
0.001371903	AVRemover	0.000553779	0.000539368
}{\mybenigntable}

\pgfplotstableread{
x              y              x-min              x-max
0.007683142	FireFox	0.001730635	0.00590558
0.002392672	FileZilla	0.00198433	0.004763791
0.002392054	PWSafe	0.00233599	0.008960579
0.002420506	MP3Gain	0.002364797	0.016215477
0.008039981	ShotCut	0.007240965	0.03342735
0.002377041	TeamViewer	0.001734131	0.02241284
0.004694459	Foobar	0.00322037	0.036700363
0.006113433	7Zip	0.004368104	0.00334437
0.003973484	TurboVNC	0.003845423	0.02166131
0.004942198	WinMerge	0.003780698	0.021243075
0.001925459	Launchy	0.001527065	0.003950908
0.004102804	Skype	0.002681407	0.024314874
0.009719952	WinRAR	0.003272973	0.016908484
0.003610377	DropBox	0.002208239	0.002505092
0.004712766	Slack	0.002649036	0.008104415
0.003706129	Flash	0.002209844	0.007435541
0.007864276	OneDrive	0.007252427	0.009834722
0.005707822	NotePad++	0.003722672	0.017186294
0.006980418	ICBC	0.006961546	0.003370005
0.003800491	AVRemover	0.001386511	0.011509088
}{\myattacktable}

\newcommand*{\benignblue}{%
   \setbox0=\hbox{\strut}%
   \begin{tikzpicture}
     \filldraw[draw=blue,fill=blue] (0,0\ht0) circle[radius=.35em];
   \end{tikzpicture}%
}

\newcommand*{\attackred}{%
   \setbox0=\hbox{\strut}%
   \begin{tikzpicture}
     \filldraw[draw=red!75,fill=red!75] (0,0\ht0) circle[radius=.35em];
   \end{tikzpicture}%
}

\begin{figure}[t]
\begin{tikzpicture}
\tikzset{mark options={mark size=0.5, opacity=0.5}}
\begin{axis} [
    xmin=0.00000005,
    xmax=0.1,
    symbolic y coords={FireFox, FileZilla, PWSafe, MP3Gain, ShotCut, TeamViewer, Foobar, 7Zip, TurboVNC, WinMerge, Launchy, Skype, WinRAR, DropBox, Slack, Flash, OneDrive, NotePad++, ICBC, AVRemover},
    xmode=log,
    tick label style={font=\scriptsize},
    xtick pos=left,
    ytick pos=left,
    ymajorgrids,
    ytick=data,
]
\addplot [only marks, blue, mark=*, mark options={blue}] 
 plot [error bars/.cd, x dir=both, x explicit]
 table [x error plus=x-max, x error minus=x-min] {\mybenigntable};
 \addplot [only marks, red!75, mark=*, mark options={red!75}] 
 plot [error bars/.cd, x dir=both, x explicit]
 table [x error plus=x-max, x error minus=x-min] {\myattacktable};
\end{axis} 
\end{tikzpicture}
\caption*{\benignblue: Benign Installer  \attackred: Malicious Installer}
\caption{Sensitivity analysis to determine the normality threshold for each software installer in the experiment. We use a log-10 scale for x-axis.}
\label{fig:eval:sensitivity}
\vspace*{-20pt}
\end{figure}

\subsection{Robustness Against Data Contamination}
\label{eval:results:contamination}
\definecolor{bblue}{HTML}{5D8AA8}
\definecolor{rred}{HTML}{E32636}
\definecolor{ggreen}{HTML}{FF9966}
\definecolor{ppurple}{HTML}{702963}
\definecolor{cadet}{HTML}{536872}
\pgfplotsset{ /pgfplots/ybar legend/.style={ /pgfplots/legend image code/.code={ \draw[##1,/tikz/.cd,bar width=3pt,yshift=-0.5em,bar shift=0pt] plot coordinates {(2*\pgfplotbarwidth,0.6em)};}, } }
\pgfplotsset{compat=1.5}

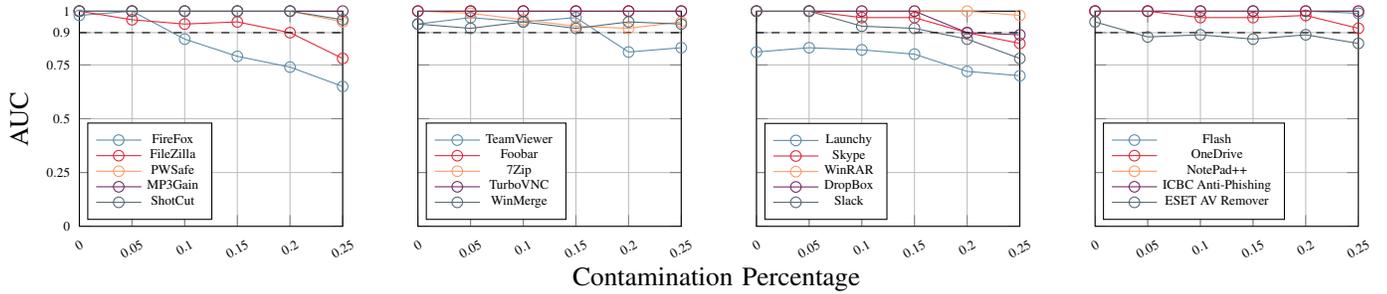
\begin{figure*}
\renewcommand\thesubfigure{(\alph{subfigure})}
	\centering
	
	\begin{tikzpicture}
	
	\begin{groupplot}[group style={
	    group size=4 by 1,
            xticklabels at=edge bottom,
            yticklabels at=edge left,
        }, 
        height=0.25\textwidth,
        width=0.6\columnwidth,
    ]
    
    \nextgroupplot[
		ylabel={AUC},
		ymin=0,
		ymax=1,
		ytick={0, 0.25, 0.5, 0.75, 0.9, 1.00},
		xmin=0,
		xmax=0.25,
		xtick={0, 0.05, 0.10, 0.15, 0.20, 0.25},
		xticklabel style={rotate=30},
		tick label style={font=\tiny},
		xticklabel style={
                 /pgf/number format/fixed, 
	         /pgf/number format/precision=2},
		grid=major,
		legend pos=south west,
		legend style={
        		font=\tiny,
		row sep=-3pt
        }
	]
   \addplot+[mark=o, mark size=2pt, color=bblue] table {
   0.00 0.98
   0.05 1.00
   0.10 0.87
   0.15 0.79
   0.20 0.74
   0.25 0.65
   };
   \addlegendentry{FireFox}
   \addplot+[mark=o, mark size=2pt, color=rred] table {
   0.00 1.00
   0.05 0.96
   0.10 0.94
   0.15 0.95
   0.20 0.90
   0.25 0.78
   };
   \addlegendentry{FileZilla}
   \addplot+[mark=o, mark size=2pt, color=ggreen] table {
   0.00 1.00
   0.05 1.00
   0.10 1.00
   0.15 1.00
   0.20 1.00
   0.25 0.95
   };
   \addlegendentry{PWSafe}
   \addplot+[mark=o, mark size=2pt, color=ppurple] table {
   0.00 1.00
   0.05 1.00
   0.10 1.00
   0.15 1.00
   0.20 1.00
   0.25 1.00
   };
   \addlegendentry{MP3Gain}
   \addplot+[mark=o, mark size=2pt, color=cadet] table {
   0.00 1.00
   0.05 1.00
   0.10 1.00
   0.15 1.00
   0.20 1.00
   0.25 0.96
   };
   \addlegendentry{ShotCut}
   \addplot[dashed, mark=none, color=black] {0.90};
   
    \nextgroupplot[
  		ymin=0,
		ymax=1,
		ytick={0, 0.25, 0.5, 0.75, 0.9, 1.00},
		xmin=0,
		xmax=0.25,
		xtick={0, 0.05, 0.10, 0.15, 0.20, 0.25},
		xticklabel style={
                 /pgf/number format/fixed, 
	         /pgf/number format/precision=2},
		xticklabel style={rotate=30},
		tick label style={font=\tiny},
		grid=major,
		legend pos=south west,
		legend style={
        		font=\tiny,
		row sep=-3pt
        }
	]
   \addplot+[mark=o, mark size=2pt, color=bblue] table {
   0.00 0.94
   0.05 0.97
   0.10 0.95
   0.15 0.97
   0.20 0.81
   0.25 0.83
   };
   \addlegendentry{TeamViewer}
   \addplot+[mark=o, mark size=2pt, color=rred] table {
   0.00 1.00
   0.05 1.00
   0.10 1.00
   0.15 1.00
   0.20 1.00
   0.25 1.00
   };
   \addlegendentry{Foobar}
   \addplot+[mark=o, mark size=2pt, color=ggreen] table {
   0.00 1.00
   0.05 0.99
   0.10 0.96
   0.15 0.93
   0.20 0.92
   0.25 0.95
   };
   \addlegendentry{7Zip}
   \addplot+[mark=o, mark size=2pt, color=ppurple] table {
   0.00 1.00
   0.05 1.00
   0.10 1.00
   0.15 1.00
   0.20 1.00
   0.25 1.00
   };
   \addlegendentry{TurboVNC}
   \addplot+[mark=o, mark size=2pt, color=cadet] table {
   0.00 0.94
   0.05 0.92
   0.10 0.95
   0.15 0.92
   0.20 0.95
   0.25 0.94
   };
   \addlegendentry{WinMerge}
   \addplot[dashed, mark=none, color=black] {0.90};
    
    \nextgroupplot[
		ymin=0,
		ymax=1,
		ytick={0, 0.25, 0.5, 0.75, 0.9, 1.00},
		xmin=0,
		xmax=0.25,
		xtick={0, 0.05, 0.10, 0.15, 0.20, 0.25},
		xticklabel style={
                 /pgf/number format/fixed, 
	         /pgf/number format/precision=2},
		xticklabel style={rotate=30},
		tick label style={font=\tiny},
		grid=major,
		legend pos=south west,
		legend style={
        		font=\tiny,
		row sep=-3pt
        }
	]
   \addplot+[mark=o, mark size=2pt, color=bblue] table {
   0.00 0.81
   0.05 0.83
   0.10 0.82
   0.15 0.80
   0.20 0.72
   0.25 0.70
   };
   \addlegendentry{Launchy}
   \addplot+[mark=o, mark size=2pt, color=rred] table {
   0.00 1.00
   0.05 1.00
   0.10 0.97
   0.15 0.97
   0.20 0.90
   0.25 0.85
   };
   \addlegendentry{Skype}
   \addplot+[mark=o, mark size=2pt, color=ggreen] table {
   0.00 1.00
   0.05 1.00
   0.10 1.00
   0.15 1.00
   0.20 1.00
   0.25 0.98
   };
   \addlegendentry{WinRAR}
   \addplot+[mark=o, mark size=2pt, color=ppurple] table {
   0.00 1.00
   0.05 1.00
   0.10 1.00
   0.15 1.00
   0.20 0.90
   0.25 0.89
   };
   \addlegendentry{DropBox}
   \addplot+[mark=o, mark size=2pt, color=cadet] table {
   0.00 1.00
   0.05 1.00
   0.10 0.93
   0.15 0.92
   0.20 0.87
   0.25 0.78
   };
   \addlegendentry{Slack}   
   \addplot[dashed, mark=none, color=black] {0.90};
    
    \nextgroupplot[
    		ymin=0,
		ymax=1,
		ytick={0, 0.25, 0.5, 0.75, 0.9, 1.00},
		xmin=0,
		xmax=0.25,
		xtick={0, 0.05, 0.10, 0.15, 0.20, 0.25},
		xticklabel style={
                 /pgf/number format/fixed, 
	         /pgf/number format/precision=2},
		xticklabel style={rotate=30},
		tick label style={font=\tiny},
		grid=major,
		legend pos=south west,
		legend style={
        		font=\tiny,
		row sep=-3pt
        }
        ]
   \addplot+[mark=o, mark size=2pt, color=bblue] table {
   0.00 1.00
   0.05 1.00
   0.10 1.00
   0.15 1.00
   0.20 1.00
   0.25 0.99
   };
   \addlegendentry{Flash}
   \addplot+[mark=o, mark size=2pt, color=rred] table {
   0.00 1.00
   0.05 1.00
   0.10 0.97
   0.15 0.97
   0.20 0.98
   0.25 0.92
   };
   \addlegendentry{OneDrive}
   \addplot+[mark=o, mark size=2pt, color=ggreen] table {
   0.00 1.00
   0.05 1.00
   0.10 1.00
   0.15 1.00
   0.20 1.00
   0.25 1.00
   };
   \addlegendentry{NotePad++}
   \addplot+[mark=o, mark size=2pt, color=ppurple] table {
   0.00 1.00
   0.05 1.00
   0.10 1.00
   0.15 1.00
   0.20 1.00
   0.25 1.00
   };
   \addlegendentry{ICBC Anti-Phishing}
   \addplot+[mark=o, mark size=2pt, color=cadet] table {
   0.00 0.95
   0.05 0.88
   0.10 0.89
   0.15 0.87
   0.20 0.89
   0.25 0.85
   };
   \addlegendentry{ESET AV Remover}
   \addplot[dashed, mark=none, color=black] {0.90};
    
    \end{groupplot}
    \node[text width=6cm,align=center,anchor=north] at (\columnwidth, -12pt) {Contamination Percentage};
    \end{tikzpicture}
    \caption[]{AUC result breakdown for each software installer with various degrees of data contamination.}
    \label{fig:contamination}
    \vspace*{-15pt}
\end{figure*}

So far, we have assumed that anomaly-free data is available for training,
but this assumption does not hold in most real-life scenarios.
On the contrary,
real-world data often contains noise or undetected anomalies
(\ie contaminations) that potentially affect detection performance~\cite{berg2019unsupervised}.
Hence, a fully unsupervised learning system requires a certain degree of robustness
that minimizes the need for weak labeling of benign data~\cite{khoshnevisan2019rsm}.
We evaluate the effects of anomaly contaminations in the training set
for each software installer in~\autoref{table:eval:datasets}.

\noindgras{Experimental Setup.}
We contaminated 5\%, 10\%, 15\%, 20\%, and 25\% of the original training set
with malware data from the test set and rebuilt the model for each level of contamination.
Malware data used for training is also included in the test set to evaluate \system's robustness 
against anomaly data pollution.
We use the Area Under the Receiver Operating Characteristics (ROC) curve, or AUC,
to compare anomaly detection results for each installer (\autoref{fig:contamination}).
AUC, ranging between 0 and 1, measures the quality of model prediction regardless of classification threshold.

\noindgras{Experimental Results.}
\autoref{fig:contamination} shows that in general, \system is tolerant to contamination in training data.
In the majority of cases,
the AUC stays above 0.90,
even when contamination is severe (\eg 25\%).
We notice that applications with lower performance in \autoref{sec:eval:results} (\eg \texttt{FireFox}) 
are more likely to be affected by contamination,
as their benign installation behavior is already difficult to learn
even with clean training data.

\subsection{Robustness Against Adversarial Attacks}
\label{eval:results:adversarial}
With the growing popularity of graph-based classification methods in security applications,
adversarial attacks on graph data are likely to become increasingly common for an attacker to evade those methods~\cite{wang2019attacking}.
However, 
there exist only a few studies~\cite{dai2018adversarial, zugner2018adversarial, wang2019attacking, zugner2019certifiable} on this topic, 
with the majority focusing on citation networks (\eg Cora~\cite{mccallum2017cora}, Citeseer~\cite{caragea2014citeseer}) and social networks
(\eg Facebook, Twitter~\cite{wang2017sybilscar}),
and designed only for a particular type of graph neural networks (\eg GCN~\cite{zugner2019certifiable}).

To demonstrate \system's robustness against adversarial attacks,
we investigate two realistic attack scenarios from a \emph{practical, systems} perspective.
Different from prior approaches that focus on network graph attacks,
our scenarios require a distinct set of attacker behavior (and thus resulting graph perturbations), 
constrained by the threat model (\autoref{sec:threat}), our neural network architecture and classification method,
but more importantly, the feasibility of system manipulations.

\noindgras{Background.}
We consider the \emph{restrict black-box attack (RBA)} and \emph{practical black-box attack (PBA)} adversarial settings~\cite{dai2018adversarial}
\footnote{We do not consider the \emph{white-box attack (WBA)} setting in which the attacker can access any model information, 
including model parameters and gradient information,
since such accessibility is rarely possible in real-life situations~\cite{changrestricted}.}.
In RBA, 
the attacker must perform adversarial graph modifications without any knowledge of our model,
given only sampled benign and attack graphs.
The PBA scenario relaxes the restrictions on model knowledge
by disclosing discrete prediction feedback from the target classifier (but not any other information \eg the normality threshold).
Our threat model assumes the integrity of data provenance,
so the attacker cannot directly modify SIGs.
They can manipulate graph structures (\ie \emph{structure attack}) and
node feature vectors (\ie \emph{feature attack}) only by
manipulating software installation process,
while ensuring successful malware execution.

We follow state-of-the-art graph-based adversarial machine learning literature~\cite{wang2019attacking, zugner2018adversarial}
to generate adversarial attack graphs by
1) adding or removing edges, and
2) modifying node attributes on the malicious graphs in \autoref{table:eval:datasets}.
As discussed in detail below,
we also define an \emph{equivalency indicator}~\cite{dai2018adversarial} for each attack setting
to restrict graph perturbations that are realistically available to the attacker
(\eg the attacker cannot add a directed edge between two file nodes).

\noindgras{Experimental Setup (RBA).}
We define the equivalency indicator as any allowed graph modifications
on nodes/edges related to the malicious processes.
The attacker can easily identify those graph components given both benign and attack graphs.
Without any additional information,
the attacker is empirically better off to focus on malicious process nodes
that typically receive high anomaly scores and influence graph classification (\autoref{sec:framework:detection}).
Conceptually, this is equivalent to adversarial attacks in node classification problems,
where malicious process nodes are the attacker's \emph{target nodes}.
Prior studies have demonstrated that manipulations on 
target nodes result in significantly more adversarial damage~\cite{zugner2018adversarial, changrestricted}. 
 
One strategy is to disguise malicious processes to mimic the benign ones.
We design a feature attack, a structure attack, and a combination of both.
In the feature attack,
we modify the malicious process' node attributes to be the same as those of the benign ones,
effectively aligning feature vectors of both malicious and benign nodes (\autoref{sec:framework:embedding}).
In the structure attack,
we ensure that the malicious processes read/write the same number of files/sockets
and fork the same number of child processes,
so that their local structures approximate those of the benign processes.
In the combination of both attacks,
we further make sure that feature vectors of files/sockets/processes related to the malicious processes
are similar to those related to the benign processes
(\eg by manipulating file node attributes).
We evaluate the effects of all attack vectors for each software installer in \autoref{table:eval:datasets}.

\definecolor{bblue}{HTML}{5D8AA8}
\definecolor{rred}{HTML}{E32636}
\definecolor{ggreen}{HTML}{FF9966}
\definecolor{ppurple}{HTML}{702963}
\pgfplotsset{compat=1.5}

\begin{figure}
\begin{tikzpicture}
\begin{axis}[
    ybar,
    legend style={at={(0.5,-0.20), font=\tiny},
      anchor=north,legend columns=-1},
    enlarge y limits={upper, value=0.3},
    ylabel={AUC},
    ymin = 0,
    symbolic x coords={FireFox, TeamViewer, WinMerge, Launchy, Slack, AV Remover},
    xtick=data,
    xtick pos=left,
    ytick pos=left,
    label style={font=\tiny},
    tick label style={font=\tiny},
    ylabel near ticks,
    xlabel near ticks,
    width=\columnwidth,
    bar width=6,
    height=0.5\columnwidth,
    nodes near coords,
    every node near coord/.append style={rotate=90, anchor=west, font=\tiny, /pgf/number format/.cd, fixed, precision=4}
    ]
    
    \addplot[fill={bblue}] coordinates {(FireFox, 0.988) (TeamViewer, 0.977) (WinMerge, 0.986) (Launchy, 0.924) (Slack, 1.0)};
    \addplot[fill={rred}] coordinates {(FireFox, 0.963) (TeamViewer, 0.969) (WinMerge, 0.961) (Launchy, 0.911) (Slack, 0.944)};
    \addplot[fill={ggreen}] coordinates {(FireFox, 0.965) (TeamViewer, 0.965) (WinMerge, 0.965) (Launchy, 0.916) (Slack, 0.942)};
    \addplot[fill={ppurple}] coordinates {(FireFox, 0.931) (TeamViewer, 0.960) (WinMerge, 0.959) (Launchy, 0.911) (Slack, 0.942)};
    \legend{No Attack, Feature Attack, Structure Attack, Combined Attack}
\end{axis}
\end{tikzpicture}
\caption{AUC result breakdown for software installers affected by RBA.}
\label{table:eval:results:blackbox}
\vspace*{-20pt}
\end{figure}
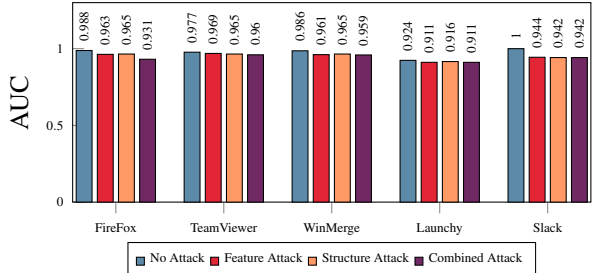

\noindgras{Experimental Results (RBA).}
\autoref{table:eval:results:blackbox} shows the results for only those
software installers affected by at least one attack vector.
AUCs of the other installers in \autoref{table:eval:datasets} remain unchanged.
We see that the efficacy of the feature and structure attack in isolation is installer independent:
while \texttt{TeamViewer} and \texttt{Slack} are slightly more vulnerable to the structure attack,
the rest are more affected by the feature attack.
Combining both feature and structure attacks improves attack performance,
but overall, \system is robust to adversarial attack in this scenario.
\system's use of deep graph learning means that changes in one part of the graph
can have far-reaching consequences.
Manipulating anomalous process nodes does not remove all the effects of such nodes;
the benign nodes to which they connect are also affected by their originally malicious behavior~\cite{zugner2018adversarial}.
The attackers could strengthen RBA if they can also accurately identify target
nodes that are \emph{not} malicious but have been influenced by the malicious processes,
but such information is not available in this setting.

\noindgras{Experimental Setup (PBA).}
PBA allows the attacker to obtain prediction feedback from the classifier,
so the attacker can iteratively add/remove edges or modify node features in the graph,
until the resulting graph produces a false negative from \system's model.
We will generate such a PBA attack using reinforcement learning (RL).
Our goal is to build an RL-model that takes as input a SIG produced
by an existing malware package and produces, as output, a SIG that \system
improperly classifies as benign.
We constrain the changes that the RL-model can make on the graph
to structural changes that can be produced according to the criteria
discussed in the previous section (\ie that the attackers can
produce manipulated graphs only by changing their attack implementation),
and define the equivalency indicator as the minimal number of such modifications within a fixed budget~\cite{zugner2018adversarial}.
We adopt a hierarchical reinforcement learning (RL) based attack method
through Q-learning to learn a \emph{generalizable} attack policy over graph structure~\cite{dai2018adversarial}.
We build our RL-model using a subset of the malware of a single application
(we randomly chose 5\% of the \texttt{Skype} malware installations) and then evaluate
the model using the full suite of malware from~\autoref{table:eval:datasets}.

\definecolor{bblue}{HTML}{5D8AA8}
\definecolor{rred}{HTML}{E32636}
\definecolor{ggreen}{HTML}{FF9966}
\definecolor{ppurple}{HTML}{702963}
\pgfplotsset{compat=1.11}

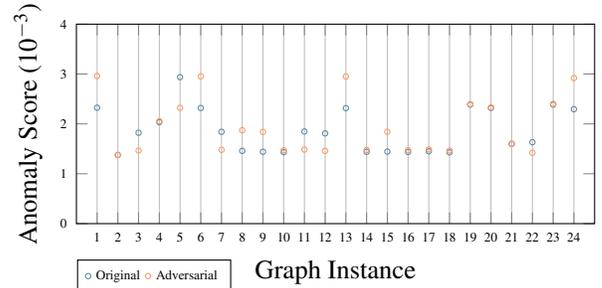
\begin{figure}
	\centering
	\begin{tikzpicture}
	\tikzset{mark options={mark size=1}}
		\begin{axis}[
		axis line style={-},
		ylabel={Anomaly Score ($10^{-3}$)},
		xlabel={Graph Instance},
		ymin=0,
		ymax=4,
		height=0.5\columnwidth,
		width=\columnwidth,
		xmin=0,
		xmax=25,
		xtick={1, 2, ..., 24},
		yticklabel style={
			/pgf/number format/fixed,
			/pgf/number format/precision=5},
		tick label style={font=\tiny},
		xmajorgrids,
		legend style={at={(0.15,-0.18), font=\tiny},
			anchor=north,legend columns=-1},
		scatter/classes={a={mark=o, bblue},
					  b={mark=o, ggreen}}
			]
	\addplot+[scatter, only marks, scatter src=explicit symbolic] table [meta=label]{
	x     y      label
	1  2.324725915818071   a
	1  2.9619022546992927  b
	2  1.3767143193981052   a
	2  1.376694017859036  b
	3  1.8222329558566321  a
	3  1.4662703213479521  b
	4  2.034822449520611  a
	4  2.0536084491742366  b
	5  2.9378113875042078  a
	5  2.3224833961988114  b
	6  2.319692869191311  a
	6  2.9531988305772847  b
	7  1.8415882287701037  a
	7  1.4801296206678246  b
	8  1.4586380991000228  a
	8  1.873102217210769  b
	9  1.4413431257407855  a
	9  1.8399283288652486  b
	10  1.4336401936320242  a
	10  1.4710571391154698  b
	11  1.8483873897200242  a
	11  1.485390303828754  b
	12  1.8093259095225264  a
	12  1.4583082198689164  b
	13  2.316332766941397  a
	13  2.9516994702460213  b
	14  1.4402274299526146  a
	14  1.476649825203646  b
	15  1.4448031567226053  a
	15  1.843606105685725  b
	16  1.436455284113683  a
	16  1.4730923333015546  b
	17  1.4502689802162346  a
	17  1.4881800356266208  b
	18  1.429307739844717  a
	18  1.4658633233512162  b
	19  2.386680795537166  a
	19  2.3959436061888727  b
	20  2.3197826667443766  a
	20  2.3332456609401176  b
	21  1.599561886071558  a
	21  1.5995501110741537  b
	22  1.6330186334328175  a
	22  1.4210430837614636  b
	23  2.3874200072030245  a
	23  2.402594007484245  b
	24  2.2951093997002864  a
	24  2.919779426108118  b
   	};
	\legend{Original, Adversarial}
		\end{axis}
	\end{tikzpicture}
\caption{Anomaly scores of \texttt{Skype} attack graphs affected by PBA.}
\label{table:eval:results:adversarial2}
\vspace*{-20pt}
\end{figure}

\noindgras{Experimental Results (PBA).}
The adversarial attacker tries to increase the false negative rate (FNR) of the attack graphs, 
but we observe no such changes for \texttt{Skype} nor for the majority of the
other installers in~\autoref{table:eval:datasets}.
The two exceptions are \texttt{TeamViewer} and \texttt{FireFox};
\texttt{TeamViewer} exhibits more FNs for one attack graph, and
\texttt{FireFox} exhibits fewer FNs for one attack graph.
When applying the adversarial model trained on the \texttt{Skype} dataset to other installers,
its performance varies depending on the installer.
In fact, its efficacy fluctuates even within the \texttt{Skype} dataset itself where the target model is known to the attacker.
We investigate the changes in anomaly scores of \texttt{Skype}'s attack graphs under the adversarial influence.
\autoref{table:eval:results:adversarial2} shows that
even the best possible manipulation (predicted by the trained RL model) does not
necessarily reduce an attack graph's anomaly score.
Our results differ significantly from
prior work demonstrating the efficacy of adversarial attacks on graphs
(e.g., ~\cite{dai2018adversarial}).
This prior work demonstrated efficacy on
graphs from citation and social networks.
We hypothesize that adversarial attacks are less effective in 
our setting, because 1) provenance graphs are structurally different from
these network graphs, and 2) our setting allows a more constrained set
of changes to the graph.

\subsection{Building \system Meta-Model}
\label{eval:results:meta}
\definecolor{bblue}{HTML}{5D8AA8}
\definecolor{rred}{HTML}{E32636}
\definecolor{ggreen}{HTML}{FF9966}
\definecolor{ppurple}{HTML}{702963}
\pgfplotsset{compat=1.5}

\begin{figure}
\resizebox{\columnwidth} {!} {
\begin{tikzpicture}
\begin{axis}[
    ybar,
    legend style={at={(0.5,-0.40), font=\tiny},
      anchor=north,legend columns=-1},
    enlarge y limits={upper, value=0.3},
    ylabel={AUC},
    ymin = 0,
    symbolic x coords={FireFox, FileZilla, PWSafe, MP3Gain, ShotCut, TeamViewer, Foobar, 7Zip, TurboVNC, WinMerge, Launchy, Skype, WinRAR, DropBox, Slack, Flash, OneDrive, NotePad++, ICBC, AV Remover},
    x tick label style={rotate=45, anchor=north east},
    xtick=data,
    xtick pos=left,
    ytick pos=left,
    label style={font=\tiny},
    tick label style={font=\tiny},
    ylabel near ticks,
    xlabel near ticks,
    width=1.4\columnwidth,
    bar width=3,
    height=0.5\columnwidth,
    nodes near coords,
    every node near coord/.append style={rotate=90, anchor=west, font=\tiny, /pgf/number format/.cd, fixed, precision=4}
    ]
    \addplot[fill={bblue}] coordinates {(FireFox, 0.988) (FileZilla, 1.0) (PWSafe, 1.0) (MP3Gain, 1.0) (ShotCut, 1.0) (TeamViewer, 0.977) (Foobar, 1.0) (7Zip, 1.0) (TurboVNC, 1.0) (WinMerge, 0.986) (Launchy, 0.924) (Skype, 1.0) (WinRAR, 1.0) (DropBox, 1.0) (Slack, 1.0) (Flash, 1.0) (OneDrive, 1.0) (NotePad++, 1.0) (ICBC, 1.0) (AV Remover, 1.0)};
    \addplot[fill={ggreen}] coordinates {(FireFox, 0.905) (FileZilla, 0.933) (PWSafe, 1.0) (MP3Gain, 1.0) (ShotCut, 1.0) (TeamViewer, 1.0) (Foobar, 1.0) (7Zip, 1.0) (TurboVNC, 1.0) (WinMerge, 0.993) (Launchy, 0.768) (Skype, 0.995) (WinRAR, 1.0) (DropBox, 1.0) (Slack, 0.960) (Flash, 0.833) (OneDrive, 1.0) (NotePad++, 1.0) (ICBC, 0.945) (AV Remover, 0.935)};
    
    \legend{Application-Specific Model, Meta-Model}
\end{axis}
\end{tikzpicture}
}
\caption{AUC result comparison for each installer using application-specific vs. meta model. The \texttt{Skype} dataset is not used in training the meta model.}
\label{table:eval:results:meta}
\vspace*{-15pt}
\end{figure}

\pgfplotsset{ /pgfplots/ybar legend/.style={ /pgfplots/legend image code/.code={ \draw[##1,/tikz/.cd,bar width=3pt,yshift=-0.5em,bar shift=0pt] plot coordinates {(2*\pgfplotbarwidth,0.6em)};}, } }
\pgfplotsset{compat=1.5}

\system is designed to build one model per application,
but it can easily build a ``meta-model'' that learns \emph{generic} software installation behavior.
Intuitively,
such a generalized model can classify unseen installers,
thus saving considerable manual labor from training new application-specific models.
On the other hand,
it must perform comparably to those models
to warrant its usability for the installers in the training dataset.

\noindgras{Experimental Setup.}
We trained a meta-model using the training sets from all but the \texttt{Skype} installer (selected randomly).
We then evaluated the meta-model using both the benign and malicious datasets from each application, including \texttt{Skype}.
This experimental setup is identical to the one described in \autoref{eval:results:setup}
to fairly compare against application-specific models.
We repeated this experiment by randomly excluding different installers;
the results are similar.

We further investigated meta-model performance when trained with various numbers of applications.
We excluded 5\%, 10\%, 20\%, and 40\% of the original applications from the training set
and rebuilt the meta-model for each scenario.
We evaluated each meta-model with two sets of test data,
1) the benign and malicious test sets from the applications used in training (INC in \autoref{table:eval:results:meta2}), and
2) the benign and malicious test sets from the excluded applications (EXC).

\definecolor{bblue}{HTML}{5D8AA8}
\definecolor{rred}{HTML}{E32636}
\definecolor{ggreen}{HTML}{FF9966}
\definecolor{ppurple}{HTML}{702963}
\pgfplotsset{compat=1.5}

\begin{figure}[t]
	\centering
	
	\begin{tikzpicture}
		\begin{axis}[
		ylabel={AUC},
		xlabel={Percentage of Excluded Applications},
		ymin=0,
		ymax=1,
		ytick={0.5, 0.75, 1.00},
		axis y discontinuity=crunch,
		height=0.4\columnwidth,
		width=\columnwidth,
		xmin=0,
		xmax=0.50,
		xtick={0, 0.05, 0.10, 0.20, 0.40, 0.50},
		xticklabel style={rotate=30},
		tick label style={font=\tiny},
		xticklabel style={
                 	/pgf/number format/fixed, 
	         	/pgf/number format/precision=2},
		grid=major,
		legend pos=south west,
		legend style={
        			font=\tiny,
			row sep=-3pt}]
	\addplot+[mark=o, mark size=2pt, color=bblue] table {
   		0.05 0.953
   		0.10 0.996
   		0.20 0.996
   		0.40 0.999
   	};
	\addlegendentry{INC}
	\addplot+[mark=o, mark size=2pt, color=ggreen] table {
   		0.05 0.995
   		0.10 1.0
   		0.20 0.954
   		0.40 0.720
   	};
	\addlegendentry{EXC}
		\end{axis}
	\end{tikzpicture}
\caption{AUC results when meta-models are trained with various numbers of applications.
The meta-models are tested on applications included (INC) in and excluded (EXC) from the training data.}
\label{table:eval:results:meta2}
\vspace*{-15pt}
\end{figure}

\noindgras{Experimental Results.}
\autoref{table:eval:results:meta} shows the AUC results for all the installers.
For half of the installers, the AUC is unchanged;
even for the other half, it decreases marginally.
Most installers achieve over 0.9 AUC under the meta-model.
Although the model is never trained on the \texttt{Skype} dataset,
it is able to accurately separate its benign and malicious instances.
This result implies that commonalities exist in legitimate software installations,
and \system learns these shared characteristics.
Surprisingly, we also see AUC \emph{improvement} for \texttt{TeamViewer} and \texttt{WinMerge},
which is likely the result of model generalizability.
\autoref{table:eval:results:meta2} shows the AUC results for meta-models
trained with different percentages of applications. 
When the meta-model learns from a smaller set of applications,
it inevitably faces more challenges generalizing to unseen software,
but works better on the trained ones.
Since the performance gracefully degrades with an increasing number of new applications,
\system provides abundant opportunities for system administrators to retrain the meta-model (~\autoref{sec:discussion}).

\subsection{Runtime Performance}
\label{eval:results:perf}


\system takes, on average, fewer than 90 minutes (on a single GPU on our local test machine) to train a model for a particular software.
Training for different installations can be performed in parallel and/or distributed to the cloud.
\autoref{table:eval:datasets} shows the number of installation graphs we used for training.
We train only on the graphs available in our current database;
\system can be effective even across versions (\autoref{eval:results:use})
and on unseen software (\autoref{eval:results:meta}).
\system supports incremental learning to efficiently train on new graph samples. 
With \system's guidance (\autoref{eval:results:investigation}),
system administrators can easily decide to further improve a model if top-ranked processes are not malicious.
Once trained,
\system takes less than $1$ second to evaluate a SIG.

\subsection{\system in Linux}
\label{eval:results:linux}
We see in~\autoref{eval:results:meta} that \system can build generic, \emph{application-agnostic} models that detect
abnormal installation behavior on Windows.
In this section,
we further demonstrate that \system is generalizable to an even larger variety of software packages 
and on different platforms.
Since our enterprise monitoring system collects only Windows audit data,
we set up our own Linux testbed and generated a dataset of 2,885 Python package installation graphs.

\noindgras{Experimental Setup.}
We trained \system on 1,708 benign installation graphs, each of which was collected using Linux Audit
from installing different Python packages 
including popular tools~\cite{pypi} such as \texttt{urlib3}, and \texttt{six}.
After training such a meta-model on all 1,708 packages, 
we design our experiments to focus on two research questions:

\noindgras{Q1.}
\emph{Given that \system is trained on a large number of distinct software packages,
is it able to generalize to new benign packages and maintain a low false positive rate (FPR)?}
We are particularly concerned with FPs, because
anomaly-based systems are generally more likely to produce excessive FPs that overwhelm cyberanalysts,
especially when they are trained on diverse datasets.
We tested the model on 1,176 installation graphs of benign packages unknown to the model.

\noindgras{Q2.}
\emph{Can \system accurately detect malicious software packages and provide targeted guidance?}
We used a real-world malicious Python package \texttt{python3-dateutil}
that was uploaded to PyPI in 2019.
The benign version of the same package is a popular utility tool that extends Python's standard \texttt{datetime} module.
We note that the attack does not create any malicious binary files on the victim system.
Instead, it executes obfuscated malicious code in the package that transmits sensitive user information to a remote host.

\noindgras{Experimental Results (Q1).}
Among 1,176 benign test graphs,
\system reports 29 FPs, resulting in only 2.47\% FPR. 
This further corroborates our experimental results in~\autoref{eval:results:meta} that
\system is capable of learning from a diverse set of training data to model generic installation behavior.

\noindgras{Experimental Results (Q2).}
\system correctly detects the malicious Python package.
It indicates the process making a network connection to a Bitly URL as the most abnormal,
thus providing accurate attack attribution.

Overall, 
\system is effective in modeling diverse installation behaviors from a large variety of software packages
on different OS platforms and installation frameworks.

\section{Case Studies}
\label{sec:case}
We describe two case studies illustrating
\system using different real-world malicious installers in~\autoref{table:eval:malinstaller}.

\noindgras{Malware Bundled with \eset Installer.}
\label{sec:case:bundle}
In~\autoref{sec:motivation},
we described a real-world attack scenario
where the user is phished to install a legitimate \eset installer~\cite{Ransom:Win32/Wadhrama.A!rsm}
bundled with malware.
\autoref{img:motivation} shows a simplified software installation graph
from this scenario.
When the malware (\texttt{taskhost.exe} in the shaded area in~\autoref{img:motivation}) runs 
during benign software installation (\texttt{AVRemover.exe}),
it establishes a communication channel (\texttt{x.y.z.s:t}) with the attacker,
which allows the attacker to perform further damage (\eg exfiltrate sensitive information).
Note that the user is unaware of this activity
since she is distracted interacting with the benign \eset installer.

We discuss in~\autoref{sec:motivation}
how existing tools might fail to detect malicious activities from such an installation.
\system, on the other hand,
constructs a SIG from the audit data,
and tests the graph against the existing \eset model.
\system generates a threat alert for this graph
because its anomaly score is much larger than the set threshold and 
orders of magnitude greater than those of the training graphs.
\system also ranks the \texttt{AVRemover.exe} process node in the shaded area in~\autoref{img:motivation}
among the most anomalous processes (\ie targeted guidance).
We observe that \texttt{AVRemover.exe} is considered more anomalous
than the malware process \texttt{taskhost.exe},
probably because
it is uncommon for the installer process to spawn two child processes at the beginning of the installation.
\system ranks the malware process \texttt{taskhost.exe} lower
because structurally, it resembles benign process behavior that also communicates with outside IP addresses.
However, 
system administrators can easily identify the malicious process
through quick one-hop backtracking starting from the top-ranked \texttt{AVRemover.exe} process.
Compared to the entire SIG,
\system reduces the number of events that the administrator needs to inspect by two orders of magnitude.

\begin{figure}[t]
	\centering
	\includegraphics[width=\columnwidth]{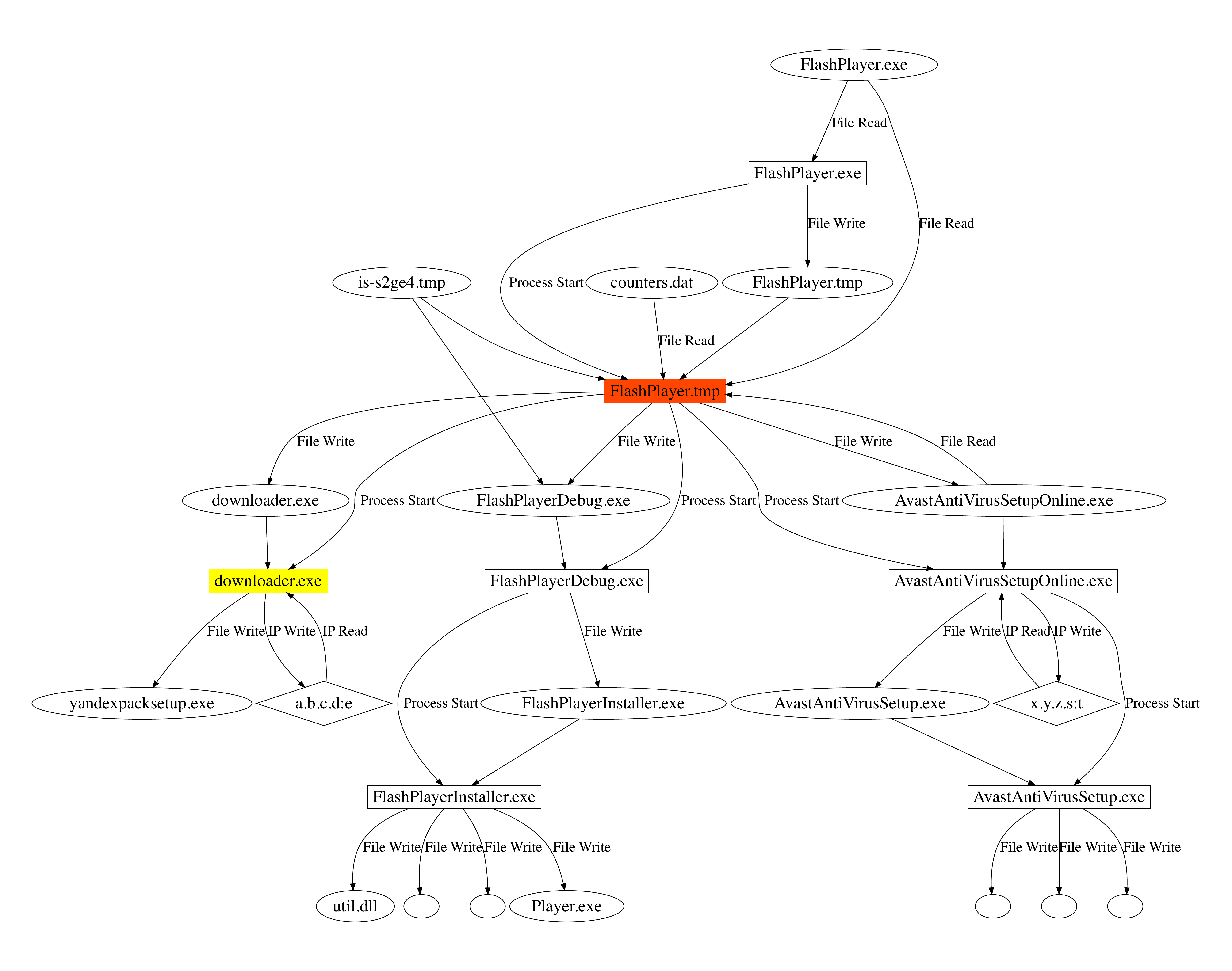}
	\caption{The software installation graph from the malicious \texttt{Flash} installer. 
	The colored process nodes are top-ranked by \system.}
	\label{img:flash}
	\vspace*{-20pt}
\end{figure}

\noindgras{Malware Embedded within \texttt{Flash} Installer.}
\label{sec:case:embed}
Different from the malicious \eset installer,
the malicious \texttt{Flash} installer embeds a dropper 
and a potentially unwanted application (PUA).
The dropper (\texttt{downloader.exe}) communicates with outside channels
and downloads additional malware (\eg\texttt{yandexsetup.exe}).
The installer also installs anti-virus software (\texttt{AvastAntiVirusSetup.exe}) without user consent.
\autoref{img:flash} shows a simplified software installation graph. 

\system identifies \texttt{FlashPlayer.tmp} (red) as the most anomalous process (\ie targeted guidance)
and \texttt{downloader.exe} (yellow) in the top 10.
The additional processes started by the installation process (\texttt{FlashPlayer.tmp}) and their progeny subgraphs
possibly lead to its high anomaly score.
The PUA, the dropper, and the malware it drops all behave differently from the benign \texttt{Flash} installer.
\system ranks the dropper process and all the malware processes (not shown in~\autoref{img:flash} for clarity)
above the PUA process,
because the PUA process behaves in a manner closer to that of the real installation process
(\texttt{FlashPlayerInstaller.exe})
than do the other malicious processes. 
We can see from~\autoref{img:flash} that their substructures resemble each other.
Regardless,
given the dropper process,
administrators already have sufficient information to confirm the malicious nature of the installation.

\section{Discussion \& Limitations}
\label{sec:discussion}

\system's ML model shares characteristics common to other statistical models~\cite{tsai2009intrusion};
model performance improves with more training data.
As we see in~\autoref{sec:eval:results} and \autoref{eval:results:comparison},
\system achieves good detection performance with only
a small number of benign installation graphs for training
because of the specificity of the domain, which enables \system
to quickly learn representative behavior patterns. 
Other deep-learning-based detection systems, 
\eg DeepLog~\cite{du2017deeplog} and Tiresias~\cite{shen2018tiresias},
also enjoy the same advantage as they target specific areas in the security domain.
For example,
DeepLog mines log data in regulated environments such as Hadoop
and thus can learn normal application behavior from a small fraction of normal log entries.

Regardless of training data size,
one important key to \system's success,
and of any modeling-based system, is data quality.
We see in~\autoref{sec:eval:results} that when data quality deteriorates,
it adversely affects system performance.
However,
\system can significantly outperform its peer systems,
even with fairly limited training data.
We attribute its efficacy to the fact that \system learns on the entire graph,
not a summary of it.
This makes \system desirable in an enterprise environment where the only training
data available have been generated internally or in which the third party tools
that collect the data might lose data, \eg due to small buffers
or slow ingestion rates~\cite{windowsetweventtracing}.

\noindgras{Software Evolution.}
We see that \system delivers consistent performance across software versions (\autoref{eval:results:use}) and 
builds application-agnostic models with a diverse training dataset (\autoref{eval:results:meta}, \autoref{eval:results:linux}).
It can also learn deltas of software versions,
by modeling past versions of software, which we leave for future work.
However, as software continues to evolve and additional software packages are installed,
\system may eventually require retraining on the SIGs of new installers.
%
%
We lessen such burdens in several ways:
1) \system maintains a good margin between anomaly scores of benign and malicious installers (\autoref{eval:results:sensitivity}).
	System administrators can easily position an installer's anomaly score among those used in training
	and determine whether retraining is necessary.
	For example, 
	the benign \texttt{NotePad++} installer with the highest anomaly score 
	($1.233\times10^{-4}$)
	is, in fact, the older version,
	while the training instances used to model the newer version have much lower scores (between $5\times10^{-6}$ and $5\times10^{-5}$).
	Admins might want to consider retraining if they want all benign instances to have anomaly scores $ < 1\times10^{-4}$.
	%
2) \system provides effective guidance (\autoref{eval:results:investigation}) to help analysts 
	identify alert causes and dismiss false positives.
	%
3) \system's performance degrades slowly (\autoref{eval:results:meta}).
	%
4) \system's retraining is fast (\autoref{eval:results:perf}).

\noindgras{Evasion.}
Stealthy malware might leverage process injection techniques
(\eg DLL injection~\cite{naikon}) to inject malicious code into a legitimate live process.
If SIG did not capture the causality relationship between the malware
and the legitimate process as a result of the injection,
the attacker could evade detection. This may be the case
given that our current prototype monitors only a subset of system events,
but state-of-the-art provenance-capture systems~\cite{pasquier2017practical} are capable of
tracking memory-related events between processes, which would allow
\system to include affected legitimate processes into analysis.
We leave as future work 
to show that
such evasion is a mere artifact of our prototype, not the approach.

Attackers might use software installation to deposit malicious software on
a system but delay exploiting that software.
As \system is optimized for detecting malicious installations, such
a deployment might go unnoticed: \system might notice that an
extra piece of software appeared, but if that software is not executed
during the installation process, \system might not flag its existence as
an anomaly.
One possible solution is to leverage forward tracking~\cite{king2005enriching}
to obtain a broader view of system behavior to detect such time-dispersed anomalies.
Prior work~\cite{milajerdi2019holmes} has shown that 
data provenance facilitates such analysis by closely connecting causal events,
even if they are temporally distant.
This makes it manageable to incorporate forward tracking into \system. 
Interesting future work would quantify the amount of 
tracking necessary for detection.

\noindgras{Benign Dataset.}
Many enterprises tightly control software installation via centralized IT departments.
Best practices for deploying new software are to test initially on a limited set of canary machines to detect
stability or compatibility issues; those machines are a natural source of labeled installation data.
Our IT department at \nec also places remote telemetry facilities on end-user machines,
collecting data using enterprise-wide security monitoring solutions.
Although we cannot guarantee the collected data is perfectly clean; in practice,
our evaluation in~\autoref{eval:results:contamination} demonstrates that \system is robust against potential data contamination.

\noindgras{Adversarial Robustness.}
We evaluated two realistic adversarial scenarios in~\autoref{eval:results:adversarial},
considering systems constraints that are absent in existing ML literature.
We show that \system is robust against practical adversarial attacks,
which is consistent with recent studies~\cite{dai2018adversarial, zugner2018adversarial}
showing that effectively attacking graph structured data is hard.
Granted, our evaluation is by no means complete
given increasing interests in ML to advance 
graph-based adversarial attacks.
For example, 
Chang \etal~\cite{changrestricted} recently proposed a graph signal-processing-based approach
to attack the graph filter of given models, 
nullifying the need for any model information.
Dai \etal~\cite{dai2018adversarial} proposed a genetic-algorithm-based attack in PBA
(although it requires additional information, \eg a normality threshold).
However,
these approaches are evaluated on the same citation network datasets,
which are structurally different from provenance graphs (\autoref{eval:results:adversarial}).
Further technical discussion and evaluation of adversarial ML is 
beyond the scope of this paper.

\section{Related Work}
\label{sec:rw}

Traditional approaches to securing software installations
emphasize authentication~\cite{bellissimo2006secure} (\eg code signing~\cite{samuel2010survivable} and secure content distribution~\cite{misra2013secure}),
policy-guided sandboxing~\cite{xu2011detecting}, 
and information flow control (IFC)~\cite{sze2015provenance}.
Recent incidents~\cite{shadowhammer, twist2017cyber} show that
attackers can compromise legitimate software distribution channels,
bypassing cryptographic authentication protection.
Meanwhile, in an enterprise environment,
sandboxing becomes impractical
and is routinely bypassed through social engineering and advanced exploit techniques~\cite{hossain2017sleuth};
sophisticated policy-driven IFC is still too complex to be widely adopted~\cite{wang2019riverbed}.
\system leverages audit data easily collectable from enterprise workstations.
Its core design lies at the intersection of graph-based malware detection
and provenance-based intrusion detection.
We place \system in the context of prior work in these areas.

\noindgras{Graph-Based Malware Detection.}
\label{sec:rw:graph}
Panorama~\cite{yin2007panorama} uses taint graphs to detect
privacy-breaching malware.
It analyzes information access and processing behavior of software
to identify violations of policies that indicate suspicious behavior traits.
Panorama generalizes signature-based malware detection
to a behavior problem like \system does,
but ultimately requires a ``behavior-signature'' that limits its detection scope.

Polonium~\cite{nachenberg2010polonium} and Marmite~\cite{stringhini2017marmite}
detect malware through large-scale graph mining
on a machine-file graph. They compute file reputation scores
and identify malware as files with low reputation.
Mastino~\cite{rahbarinia2016real} improves upon Polonium
and introduces additional URL nodes 
to graph analysis, training classifiers for URLs and files.
These approaches require 
network- and system-level data from machines across the Internet,
which is unattainable in a typical enterprise.
They consider relationships between users (\eg machines)
and files only, 
assuming that malicious files appear on few machines
and on machines with low reputation.
Such assumptions however, are no longer valid
as recent supply chain attacks leverage legitimate channels
to distribute malware to a large number of victim machines.
Kwon \etal~\cite{kwon2015dropper} proposed a downloader-graph abstraction
that describes relationships between downloaders and payloads on 5 million end-point workstations.
Using hand-crafted graph features as strong indicators of malicious activity,
the authors constructed a random forest model for malware detection.
The approach however, requires a large amount of data
(\eg features from about 24 million distinct files)
to achieve high accuracy
and any changes in malware delivery mechanisms that affect those cherry-picked features
are likely to invalidate the model.

Many other graph-based malware detection approaches exist,
with the majority focusing on characterizing malware delivery networks
~\cite{stringhini2013shady, invernizzi2014nazca}.
We omit discussions of those approaches 
since \system targets local end-point protection \emph{without} knowledge of global malware networks.
\system does not rely on extracting indicators
that signify typical cybercriminal operations,
but learns to generalize expected behavior of a particular enterprise 
given easily-accessible audit information.
Nevertheless,
a security-aware enterprise should leverage both global and local information,
complementing \system with existing global malware network analytic tools.

\noindgras{Provenance-Based Intrusion Detection.}
\label{sec:rw:anomaly}
Frappuccino~\cite{han2017frappuccino} analyzes system-level provenance graphs
to model the behavior of Platform-as-a-Service applications.
It uses a dynamic sliding window algorithm to continuously monitor and check 
if application instances conform to the learned model.
StreamSpot~\cite{manzoor2016fast} uses a similar analytic framework.
Both systems featurize provenance graphs using a \emph{bag-of-subtrees} approach
and apply clustering algorithms to identify outlier graphs.
Compared to \system's graph LSTM architecture,
learning graphs using bag-of-subtrees is insufficient to capture
the semantics of system evolution represented in provenance graphs,
due to its insensitivity to the event order.
This limitation (\ie order-insensitivity) is well-understood in NLP~\cite{tai2015improved}
and equally applicable in our domain.
Clustering bag-of-subtrees is a reasonable step
to perform outlier detection,
but it 
burdens cyberanalysts with labor-intensive investigation,
because even a single outlier often 
entails investigating a large provenance (sub)graph. 
\system lessens such a burden by triaging 
abnormal process nodes within the graph.

Recently,
Han~\etal~\cite{han2020unicorn} designed a realtime anomaly detection
system that analyzes streaming provenance graphs generated from system activity.
It learns a dynamic execution model as the host system evolves, thus capturing behavioral changes in the model.
This learning approach makes it suitable for detecting long-running persistent threats.
Gao~\etal~\cite{gao2018saql} designed a domain-specific query language, SAQL, to analyze large-scale provenance data
and use various 
anomaly models to detect intrusions.
To our best knowledge,
\system is the first provenance-based anomaly detection system
that secures software installations without prior attack knowledge.

\section{Conclusion}
\label{sec:conclusion}
We present \system,
a malware detection system that secures software installation
by analyzing the behavior of end-point systems through software installation graphs.
\system uses a novel deep graph learning architecture 
to understand installation behavior and assist attack attribution.
Our evaluation results show that \system
achieves high detection performance using only a small amount of training data,
while accurately guiding human analysts to identify the cause of alarms.
\system is therefore a practical tool that can be deployed in any enterprise
for effective and labor-saving malware detection.

\section*{Acknowledgments}
\noindent We thank the anonymous reviewers and our shepherd Konrad Rieck who helped improve the paper.
This research was supported in part by the US National Science Foundation under grant NSF 14-50277.
We acknowledge the support of the Natural Sciences and Engineering Research Council of Canada (NSERC).
Cette recherche a \'et\'e financ\'ee par le Conseil de recherches en sciences naturelles et en g\'enie du Canada (CRSNG).
The views, opinions, and/or findings contained in this paper are those of the authors and should not be interpreted as representing the official views or policies, either expressed or implied, of the sponsors.

\bibliographystyle{acm}
\footnotesize
\bibliography{biblio,bates-bib-master}


\end{document}